\documentclass[12pt,letterpaper]{article}
\pdfoutput=1
\usepackage{graphicx,array}
\usepackage{color}
\usepackage{latexsym}
\usepackage{amsthm}
\usepackage{amsmath}
\usepackage{amssymb}

\usepackage{float}
\usepackage{url}
\usepackage{mathrsfs}
\usepackage{caption}

\usepackage{dsfont}
\usepackage{epsfig}
\usepackage{slashed}
\usepackage{bbold}
\usepackage{psfrag}
\usepackage[svgnames]{xcolor}
\PassOptionsToPackage{caption=false}{subfig}
\usepackage{subcaption}
\usepackage{xfrac}
\usepackage{multirow}
\usepackage{booktabs}
\usepackage[colorlinks=true,linkcolor=Black,citecolor=Green,urlcolor=violet]{hyperref}
\usepackage{cite}
\usepackage[normalem]{ulem}

\newcommand{\be}{\begin{equation}}
\newcommand{\ee}{\end{equation}}
\newcommand{\bea}{\begin{eqnarray}}
\newcommand{\eea}{\end{eqnarray}}

\newcommand{\rmZ}{{\textrm{Z}}}
\newcommand{\rmW}{{\textrm{W}}}

\newcommand{\lambdacut}{\Lambda_{\textrm{cut}}}

\newcommand{\nn}{\nonumber}
\def\({\left(}
\def\){\right)}

\usepackage{soul}

\begin{document}
\begin{flushright}
CERN-TH-2017-125
\end{flushright}
\vspace{.6cm}
\begin{center}
{\LARGE \bf  Precision Probes of QCD at High Energies}
\bigskip\vspace{1cm}{

Simone Alioli$^1$, Marco Farina$^2$, Duccio Pappadopulo$^3$, and Joshua T. Ruderman$^3$ }
\\[7mm]
 {\it \small
 $^1$ CERN Theory Division, CH-1211, Geneva 23, Switzerland\\
$^2$New High Energy Theory Center, Department of Physics, Rutgers University,\\
  136 Frelinghuisen Road, Piscataway, NJ 08854, USA \\
$^3$Center for Cosmology and Particle Physics, Department of Physics, \\ 
New York University, New York, NY 10003, USA\\
 }

\end{center}

\bigskip \bigskip \bigskip \bigskip

%%%%%%%%%%%%%%%%%%%%%%%%%%%%%%%%%%%%%%%%%%%%%%%%%%%%%%%%%%%%%%%%%%%%%%%%%%
\vspace{-1.0cm}\centerline{\bf Abstract} 
\begin{quote}
New physics, that is too heavy to be produced directly, can leave measurable imprints on the tails of kinematic distributions at the LHC\@.  We use energetic QCD processes to perform novel measurements of the Standard Model (SM) Effective Field Theory.  We show that the dijet invariant mass spectrum, and the inclusive jet transverse momentum spectrum, are sensitive to a dimension 6 operator that modifies the gluon propagator at high energies.  The dominant effect is constructive or destructive interference with SM jet production.  We compare differential next-to-leading order predictions from POWHEG to public 7 TeV jet data, including scale, PDF, and experimental uncertainties and their respective correlations.  We constrain a New Physics (NP) scale of 3.5\,TeV with current data.  We project the reach of future 13 and 100 TeV measurements, which we estimate to be sensitive to NP scales of 8 and 60\,TeV, respectively.  As an application, we apply our bounds to constrain heavy vector octet colorons that couple to the QCD current.  We project that effective operators will surpass bump hunts, in terms of coloron mass reach, even for sequential couplings.
\end{quote}

%%%%%%%%%%%%%%%%%%%%%%%%%%%%%%%%%%%%%%%%%%%%%%%%%%%%%%%%%%%%%%%%%%%%%%%%%%
\vfill
\noindent\line(1,0){188}
{{\scriptsize{ \\ \texttt{$^1$ \href{mailto:simone.alioli@cern.ch}{simone.alioli@cern.ch}\\ \texttt{$^2$ \href{mailto:farina.phys@gmail.com}{farina.phys@gmail.com}\\ $^3$ \href{mailto:duccio.pappadopulo@gmail.com}{duccio.pappadopulo@gmail.com}\\ $^4$ \href{mailto:ruderman@nyu.edu}{ruderman@nyu.edu}}}}}
\newpage

%%%%%%%%%%%%%%%%%%%%%%%%%%%%%%%%%%%%%%%%%%%%%%%%%%%%%%%%%%%%%%%%%%%%%%%%%%%%%%%%%%%%%%%%%%%%%%%%%%%%%%%%%%%%%%%%%%%%%%%%%%%%%%%%%%%%%%%
\tableofcontents
\newpage
%%%%%%%%%%%%%%%%%%%%%%%%%%%%%%%%%%%%%%%%%%%%%%%%%%%%%%%%%%%%%%%%%%%%%%%%%%%%%%%%%%%%%%%%%%%%%%%%%%%
%%%%%%%%%%%%%%%%%%%%%%%%%%%%%%%%%%%%%%%%%%%%%%%%%%%%%%%%%%%%%%%%%%%%%%%%%%
\section{Introduction}
\label{sec:intro}
%%%%%%%%%%%%%%%%%%%%%%%%%%%%%%%%%%%%%%%%%%%%%%%%%%%%%%%%%%%%%%%%%%%%%%%%%%

The hierarchy problem of the Standard Model (SM) singles out the TeV scale as special. New particles and dynamics are expected to appear around this scale, in order to allow for a natural explanation of the small value of the Higgs boson mass. Ongoing searches for New Physics (NP) at the Large Hadron Collider (LHC) are targeting the TeV energy scale. 
The LHC is operating at 13 TeV (at or close to its design energy), and although much more data will be collected, the reach on NP, expressed in terms of the mass $M$ of particles that can be directly produced, is beginning to asymptote.  If there is NP near the TeV scale, we are confronted with the possibility that it has mass too heavy for direct production at the LHC\@.

If NP is too heavy to be directly produced, it can nevertheless leave traces at low energies through virtual effects. When an experiment is performed at an energy $E\ll M$, the effects of NP can be described in a completely general way in terms of the SM Effective Field Theory (EFT)~\cite{Buchmuller:1982ye,Feruglio:1992wf,Hagiwara:1993ck,Han:2004az,Giudice:2007fh,Grinstein:2007iv,Grzadkowski:2010es,Alonso:2012px,Contino:2013kra,Jenkins:2013zja,Jenkins:2013wua,Alonso:2013hga,Henning:2014wua,Falkowski:2015wza,Wells:2015uba,Henning:2015alf}
\be\label{smeft}
\mathscr L_{SM\, EFT}= \mathscr L_{SM}+\sum_i \frac{c_i}{M^2}\mathcal O_i^{(6)}+\ldots\,. 
\ee
Here one includes all operators of dimension $d$ bigger than 4 that can be constructed out of SM fields.

The value of an observable $O(E)$, characterized by some typical energy $E$, will receive corrections with respect to its SM value $O_{SM}(E)$ which can be written schematically as
\be\label{obseft}
O(E)= O_{SM}(E)\left(1+\sum_i  \kappa_i'\times\frac{c_i m^2_{SM}}{M^2}+\sum_i \kappa_i \times\frac{c_i E^2}{M^2}+\ldots\right) \, ,
\ee
where $m_{SM}$ is a typical SM mass scale and the dots represent terms further suppressed by inverse powers of $M$. $\kappa$ and $\kappa'$ are model dependent coefficients which are typically expected to be of order 1. Notice that the corrections shown in Eq.~\ref{obseft} arise from the interference between the SM and the the NP and, if $\kappa,\kappa'\neq 0$, are the leading corrections to the SM calculation.

If one is able to perform a measurement of the observable $O(E)$ with a precision \mbox{$\Delta O/O\equiv \delta<1$} then the combination $M/\sqrt c_i$ can be constrained with some accuracy. Let us consider two separate cases depending on the value of $E$:
\begin{itemize}
\item $E\sim m_{SM}$. The energy used to probe the operator is of the order of some SM mass scale. This is the case in which one tries to constrain the NP measuring the branching ratio of known SM particles, for instance the Higgs boson or a $B$-meson. In this case the two terms in Eq.~\ref{obseft} are of the same size and one is sensitive to $M \lesssim M_{max}$ with
\be
\frac{M_{max}}{\sqrt c_i}\sim\frac{m_{SM}}{\sqrt{\delta}}.
\ee
\item $E\gg m_{SM}$. This is the case in which an operator in Eq.~\ref{smeft} is probed at high energies, for instance looking at its effect on the final state kinematical distribution for some SM process. In this case, assuming $\kappa\sim 1$, one is sensitive to $M$ until 
\be
\frac{M_{max}}{\sqrt c_i}\sim\frac{E}{\sqrt\delta}.
\ee
\end{itemize}

Given a fixed total accuracy $\delta$, probing an observable at higher energy enhances the effect of the operators in Eq.~\ref{smeft} and allows to probe heavier NP\@.
The smallest value of $\delta$ that can be achieved at the LHC for a given observable is limited by systematic uncertainties of both experimental and theoretical origins.  However, the LHC is able to probe a variety of SM processes at very high energies and with very large luminosity. One then expects that, in certain cases, the accuracy gain associated to larger energy will be able to overcome the limited precision associated to $\delta $. Studies along these lines have been performed already by various collaborations~\cite{Domenech:2012ai,Contino:2013gna,Davidson:2013fxa,Biekoetter:2014jwa,deVries:2014apa,Azatov:2015oxa,Dror:2015nkp,Butter:2016cvz,Falkowski:2016cxu,Raj:2016aky,Greljo:2017vvb}. In particular, Ref.~\cite{Farina:2016rws}  has shown that already with existing 8\,TeV LHC data, neutral and charged current Drell-Yan measurements are able to set world leading constraints on certain electroweak observables previously studied at LEP\@. 

In this paper we apply this philosophy to the study of QCD at high energies. We are motivated by the existence of publicly available experimental measurements of both the dijet invariant mass distribution and the inclusive transverse momentum distribution performed by both ATLAS and  CMS at 7\,TeV center of mass energy \cite{Aad:2013tea,Aad:2014vwa,Chatrchyan:2012bja}.  Although other measurements of jet cross sections have been published by the experimental collaborations, even at higher collider energies, we found that only the aforementioned publications provide all the necessary statistical information, in particular the full set of correlated systematic uncertainties, which is needed to perform realistic fits to the data. Both inclusive jet and dijet cross sections have been studied as a probe of contact operators at the Tevatron~\cite{Cho:1993eu,Chivukula:1996yr,Dixon:1993xd}. By the previous argument, the LHC has a clear advantage over the Tevatron  due to its higher center of mass energy. 

We focus our attention on a specific dimension 6 operator\cite{Barbieri:2004qk}
\be\label{LZ}
\Delta \mathscr L= -\frac{\rmZ}{2 m_W^2}(D_\mu G^{\mu\nu A})^2.
\ee
We normalize the size of the operator to the $W$ boson mass, and define its strength with the dimensionless coefficient $\rmZ$\@. 
In addition to being a reasonable benchmark to perform our analysis, the operator in Eq.~\ref{LZ} can be understood as an oblique observable for the QCD sector of the SM \cite{Barbieri:2004qk}. This operator is generated in a variety of BSM scenarios. As a weakly coupled example, in Section~\ref{sec:vec} we consider a new heavy vector boson, a color octet under strong interactions, coupled universally to quarks through the $SU(3)_c$ gauge current. In a strongly coupled theory where the gluon is composite at a mass scale $\Lambda_c$ we also expect $\rmZ$ to be generated with a size $\rmZ\sim m_W^2/\Lambda_c^2$.

Assuming that CP is conserved by the NP, there is only one additional dimension 6 operator in the SM EFT that can be written in terms of the gluon fields only
\be\label{GGGop}
g_s f^{ABC} G_\mu^{\nu A}G_\nu^{\rho B}G_\rho^{\mu C}
\ee
This operator, however, does not interfere with any $2\to 2$ tree-level SM amplitude~\cite{Azatov:2016sqh} and we do not discuss its effects in this paper.\footnote{In~\cite{Barbieri:2004qk} the $\rmZ$ parameter was defined through
\be
\Delta \mathscr L= -\frac{\rmZ}{4 m_W^2}(D_\mu G_{\nu\rho}^A)^2.
\ee
Using the Bianchi identity for the gluon field, $D_\mu G^A_{\nu\rho}+D_\rho G^A_{\mu\nu}+D_\nu G^A_{\rho\mu}=0$ and integration by parts we find
\be\label{LZ1}
(D_\mu G_{\nu\rho}^A)^2= 2(D_\mu G^{\mu\nu A})^2-2 g_s f^{ABC} G_\mu^{\nu A}G_\nu^{\rho B}G_\rho^{\mu C}.
\ee
Eq.~\ref{LZ} is thus equivalent to the definition of~\cite{Barbieri:2004qk} up to the inclusion of the $GGG$ term.
}

%%%%%%%%%%%%%%%%%%%%%%%%%%%%%%%%%%%%%%%%%%%%%%%%%%%%%%%%%%%%%%%%%%%%%%%%%%
%%%%%%%%%%%%%%%%%%%%%%%%%%%%%%%%%%%%%%%%%%%%%%%%%%%%%%%%%%%%%%%%%%%%%%%%%%
\begin{table}[t]
\begin{center}
{\small
\begin{tabular}{c|ccc} 
Process & $A_{SM}$& $A^{(1)}$& $A^{(2)}$\\ \hline \hline \\[-1em]
$qq\to qq$ & $\frac{4}{9}\left(\frac{s^2+u^2}{t^2}+\frac{s^2+t^2}{u^2}-\frac{2}{3}\frac{s^2}{tu}\right)$& $ -\frac{16}{27}\left(\frac{s^2+3/2 t^2}{ m_W^2\, u}+\frac{s^2+3/2 u^2}{m_W^2\, t}\right)$ & $\frac{16}{27} \frac{s^2+3/4\,t^2+3/4\,u^2}{m_W^4}$\\ \\[-1em]
$qq'\to qq'$ &$\frac{4}{9}\frac{s^2+u^2}{t^2}$& $ -\frac{8}{9}\frac{s^2+u^2}{m_W^2\, t}$ & $\frac{4}{9}\frac{s^2+u^2}{m_W^4}$\\  \\[-1em]
$q\bar q\to q\bar q$ &$\frac{4}{9}\left(\frac{u^2+s^2}{t^2}+\frac{u^2+t^2}{s^2}-\frac{2}{3}\frac{u^2}{st}\right)$& $-\frac{16}{27}\left(\frac{u^2+3/2 t^2}{ m_W^2\, s}+\frac{u^2+3/2 s^2}{m_W^2\, t}\right)$ & $\frac{16}{27} \frac{u^2+3/4\,s^2+3/4\,t^2}{m_W^4}$\\  \\[-1em]
$q\bar q'\to q\bar q'$&$ \frac{4}{9}\frac{s^2+u^2}{t^2}$ & $  -\frac{8}{9}\frac{s^2+u^2}{m_W^2\, t}$ & $ \frac{4}{9}\frac{s^2+u^2}{m_W^4}$\\  \\[-1em]
$q\bar q\to q'\bar q'$&$ \frac{4}{9}\frac{t^2+u^2}{s^2}$ & $ -\frac{8}{9}\frac{t^2+u^2}{m_W^2\, s}$ & $ \frac{4}{9}\frac{t^2+u^2}{m_W^4}$\\
\end{tabular}
} 
\caption{\label{tab:M2Z}\footnotesize Contribution of Eq.~\ref{LZ} to $2\to2$ quark amplitudes. We write $\overline{|\mathcal M|^2}=g_s^4(A_{SM}+A^{(1)}\rmZ+A^{(2)}\rmZ^2)$, where $\overline{|\mathcal M|^2}$ is the matrix element square summed over final state and averaged over initial ones. $A_{SM}$ is the SM value of the matrix element square. No sum over final state quark flavor is included for $q\bar q\to q'\bar q'$. For the process $p_1p_2\to p_3p_4$ the Mandelstam variables are defined as usual as $s=(p_1+p_2)^2$, $t=(p_1-p_3)^2$, and  $u=(p_1-p_4)^2$.}
\end{center}
\end{table}
%%%%%%%%%%%%%%%%%%%%%%%%%%%%%%%%%%%%%%%%%%%%%%%%%%%%%%%%%%%%%%%%%%%%%%%%%%
%%%%%%%%%%%%%%%%%%%%%%%%%%%%%%%%%%%%%%%%%%%%%%%%%%%%%%%%%%%%%%%%%%%%%%%%%%

A simple way to understand the effect of Eq.~\ref{LZ} on jet processes at the LHC is to notice that it induces a higher derivative correction to the gluon kinetic term, so that the transverse part of the gluon propagator, $P_G^T$, is modified by a constant term proportional to $\rmZ$
\be\label{gluonPZ}
P^T_G(q^2)=\frac{1}{q^2}-\frac{\rmZ}{m_W^2}+O(\rmZ^2).
\ee
The modified propagator will affect, at leading order, the matrix element squared of $2\to 2$ quark processes as shown in Table~\ref{tab:M2Z}. While at fixed angle the SM matrix element asymptotes to a constant at high energy, the $\rmZ$ dependent terms introduce an energy growing behavior. This will be reflected in an anomalous high energy behavior of kinematical distributions for $pp\to$\,jets at the LHC\@.

An alternative way to reach the same conclusions is to use the equation of motion of the gluon field
\be\label{eom}
D_\mu G^{\mu\nu A}=-g_s\sum_q \bar q\gamma^\nu T^A q+ O(\rmZ)\equiv-g_s J^{\nu A}+O(\rmZ).
\ee
to rewrite the operator in Eq.~\ref{LZ} in terms of four fermion contact operators
\be\label{Z4f}
-\frac{\rmZ}{2 m_W^2}(D_\mu G^{\mu\nu A})^2=-\frac{\rmZ g^2_s}{2 m_W^2}J_\mu^A J^{\mu A}+O(\rmZ^2).
\ee
It follows that at LO Eq.~\ref{LZ} will only contribute to the cross section of $2\to 2$ processes with external quarks.\footnote{A third way to reach again the same conclusion is to perform the following field redefinition at order $\rmZ$
\be
G_\mu^A\to G_\mu^A +\frac{\rmZ}{2 m_W^2}D^\nu G_{\nu\mu}^A-\frac{g_s \rmZ}{2 m_W^2}J_\mu\,.
\ee
}

LHC jet physics has already been used to constrain the SM EFT, by both theorists~\cite{Cho:1993eu,Chivukula:1996yr,Dixon:1993xd,Franceschini:2011wr,Domenech:2012ai,Krauss:2016ely} and the experimental collaborations, ATLAS~\cite{Collaboration:2010eza,Aad:2012bsa,ATLAS:2012pu,Aad:2014aqa,Aad:2015eha,ATLAS:2015nsi,Aaboud:2017yvp} and CMS~\cite{Chatrchyan:2012bf,Aad:2011aj,Chatrchyan:2013izb,Khachatryan:2014cja,Sirunyan:2017ygf}. On the theory side, fully-exclusive NLO QCD predictions for jet production have been available for some time~\cite{Ellis:1985er,Ellis:1986bv,Kunszt:1992tn}. Very recently, the NNLO leading-color predictions were  presented for the single inclusive jet transverse momentum and dijet invariant mass~\cite{Currie:2016bfm,Currie:2017eqf}. In this study, we improve on previous theory analyses by using state of the art fully differential NLO predictions interfaced to parton shower, by using the most recent public data, and crucially by including all significant sources of uncertainties both of experimental and theoretical origin. To the best of our knowledge, the operator in Eq.~\ref{LZ} has not been previously considered by the experimental collaborations.

The remainder of the paper is organized as follows. In Section~\ref{sec:bounds1}, we use the 7\,TeV public analyses from both CMS and ATLAS to constrain $\rmZ$. In Section~\ref{sec:bounds2}, we project the bounds on $\rmZ$ to higher energies and luminosities, in particular we consider the case of 8 and 13\,TeV LHC and a possible circular $pp$ collider running at a center of mass energy of 100\,TeV\@. In Section~\ref{sec:eft}, we discuss the validity of our bounds within the EFT framework. In Section~\ref{sec:vec}, we introduce a simple UV completion of Eq.~\ref{LZ} in terms of a new heavy color octet boson coupled to the SM fermions through the $SU(3)_c$ gauge current. We compare the reach of ordinary resonance searches to our bounds extracted using Eq.~\ref{LZ}. In Section~\ref{sec:otherops}, we discuss other operators affecting jet physics at high energies. Finally, we conclude in Section~\ref{sec:conclusions}. In particular Fig.~\ref{summaryplot} in Section~\ref{sec:conclusions} summarizes our main results.

%%%%%%%%%%%%%%%%%%%%%%%%%%%%%%%%%%%%%%%%%%%%%%%%%%%%%%%%%%%%%%%%%%%%%%%%%%%%%%%%%%%%%%%%%%%%%%%%%%%%%%%%%%%%%%%%%%%%%%%%%%%%%%%%%%%%%%%%%%%%%%%%%%%%%%%%%%%%%%%%%%%%%%%%%%%%%%%%%%%%%%%%%%%%%%%%%%%%%%%%%%%%%%%%%%%%%%%%%%%%%%%%%%%%%%%%%%
%%%%%%%%%%%%%%%%%%%%%%%%%%%%%%%%%%%%%%%%%%%%%%%%%%%%%%%%%%%%%%%%%%%%%%%%%%
\section{Bounds from existing searches}
\label{sec:bounds1}
%%%%%%%%%%%%%%%%%%%%%%%%%%%%%%%%%%%%%%%%%%%%%%%%%%%%%%%%%%%%%%%%%%%%%%%%%

To constrain the operator in Eq.~\ref{LZ} we use the existing double differential measurements of the dijet cross section and inclusive jet cross section performed by both ATLAS and CMS collaborations at a center of mass energy of 7\,TeV~\cite{Aad:2013tea, Aad:2014vwa, Chatrchyan:2012bja}.
As anticipated in the previous Section, what makes these older analyses the most relevant to us is the fact that both collaborations provide the full statistical information which is needed to use these measurements to constrain NP\@. This information includes, in particular, the full statistical covariance matrix for the correlated systematic uncertainties.\footnote{\cite{Khachatryan:2016wdh} reports the measurement of the double-differential inclusive jet cross section at 13\,TeV with an integrated luminosity of 71\,${\textrm{pb}}^{-1}$. Even though the full statistical information is available there are not enough data to improve over the 7\,TeV measurements we study.}
The experimental analysis are briefly summarized in Table~\ref{tab:searches}. 

In order to compare with the available experimental data we have to calculate, for a given experimental distribution, both the SM prediction and the corrections from Eq.~\ref{LZ}. 

The SM predictions for all the distributions include QCD effects at the Next to Leading Order (NLO) interfaced to a parton shower using the {\tt{POWHEG}} method~\cite{Nason:2004rx,Frixione:2007vw} as implemented in the {\tt{POWHEG-BOX}} program~\cite{Alioli:2010xd,Alioli:2010xa}. Showering and hadronization are performed with {\tt{Pythia6}}~\cite{Sjostrand:2006za} and {\tt{Pythia8}}~\cite{Sjostrand:2007gs}.  Jets are reconstructed using the anti-$k_T$ jet algorithm~\cite{Cacciari:2008gp}, with different radius parameters. We produced results for $R=0.4, 0.6$, and $0.7$. We concentrate on the choices $R=0.6$ for ATLAS and $R=0.7$ for CMS, as these values are the ones for which a better fit of the SM background is obtained.\footnote{Incidentally, for values of $R \sim 0.7$ there is a compensation between the effects of showering and hadronization, which tend to decrease the cross section as $\sim 1/R$ by driving radiation outside the jet cone, and the effect of the underlying-event which instead grows proportionally to the jet area $\sim R^2$. The net result is that the cross section obtained through {\tt{POWHEG}} after the addition of full showering, hadronization and underlying-event is very close to the pure fixed-order cross section only for values of  $R \sim 0.7$. Deviations as large as $20-30$\% are present already for $R \sim 0.4$} 
The theory predictions are calculated using parton distribution functions from NNPDF3.0~\cite{Ball:2014uwa}. This choice is motivated by the availability of replicas obtained fitting a reduced set of experimental measurements which do not include jet data. As it will be explained shortly, this turns out to be useful to estimate to which extent the effects of NP affecting jet measurements could be hidden by the PDF fitting procedure.
 More details about the generation of {\tt{POWHEG}} predictions can be found in Appendix~\ref{app:genPOWHEG}.}

%%%%%%%%%%%%%%%%%%%%%%%%%%%%%%%%%%%%%%%%%%%%%%%%%%%%%%%%%%%%%%%%%%%%%%%%%%
%%%%%%%%%%%%%%%%%%%%%%%%%%%%%%%%%%%%%%%%%%%%%%%%%%%%%%%%%%%%%%%%%%%%%%%%%%
\begin{table}[t]
\begin{center}
{
\small
\begin{tabular}{c|c|c|c}
7\,TeV searches & ${\textrm{lumi}}\,[{\textrm{fb}}^{-1}]$& cuts & rapidity bins\\ \hline \hline
\multirow{ 2}{*}{ATLAS dijet \cite{Aad:2013tea} } & \multirow{ 2}{*}{4.5} & $p_T^{1(2)}>100(50)$\,GeV & $i/2<y^*<(i+1)/2$ \\ & & $|y|<3$, $R=0.6$ & $i=0,\ldots,5$ \\ \hline
\multirow{ 2}{*}{ATLAS inclusive jet \cite{Aad:2014vwa} } & \multirow{ 2}{*}{4.5} & $p_T>100$\,GeV & $i/2<|y|<(i+1)/2$ \\ & & $|y|<3$, $R=0.6$ & $i=0,\ldots,5$ \\ \hline
\multirow{ 2}{*}{CMS dijet \cite{Chatrchyan:2012bja} } & \multirow{ 2}{*}{5.0} & $p_T^{1(2)}>60(30)$\,GeV  & $i/2<{\textrm{max}}|y|<(i+1)/2$ \\ & & $R=0.7$  & $i=0,\ldots,4$ \\ \hline
\multirow{ 2}{*}{CMS inclusive jet \cite{Chatrchyan:2012bja} } & \multirow{ 2}{*}{5.0} & $p_T>100$\,GeV  & $i/2<|y|<(i+1)/2$ \\ & & $R=0.7$ & $i=0,\ldots,4$ \\ 
\end{tabular}
} 
\caption{\label{tab:searches}\footnotesize Summary of the experimental searches used for our analysis. The double differential measurements are always presented in terms of an energy variable (the dijet invariant mass, $m_{jj}$, for the dijet search and the jet transverse momentum, $p_T$, for the inclusive jet one) and a rapidity variable. The variable $y^*$, used by the ATLAS dijet analysis, corresponds to $y^*\equiv|y_1-y_2|/2$, where $y_{1,2}$ are the rapidities of the two jets. Notice that $y^*=|\log\tan\hat \theta/2|$ where $\hat\theta$ is, at leading order, the scattering angle in the center of mass frame. $R$ is the jet radius.}
\end{center}
\end{table}
%%%%%%%%%%%%%%%%%%%%%%%%%%%%%%%%%%%%%%%%%%%%%%%%%%%%%%%%%%%%%%%%%%%%%%%%%%
%%%%%%%%%%%%%%%%%%%%%%%%%%%%%%%%%%%%%%%%%%%%%%%%%%%%%%%%%%%%%%%%%%%%%%%%%%

%%%%%%%%%%%%%
%%%%%%%%%%%%%
%%%%%%%%%%%%%
%%%%%%%%%%%%%
\begin{figure}[t]
\begin{center}
~~\includegraphics[width=0.45\textwidth]{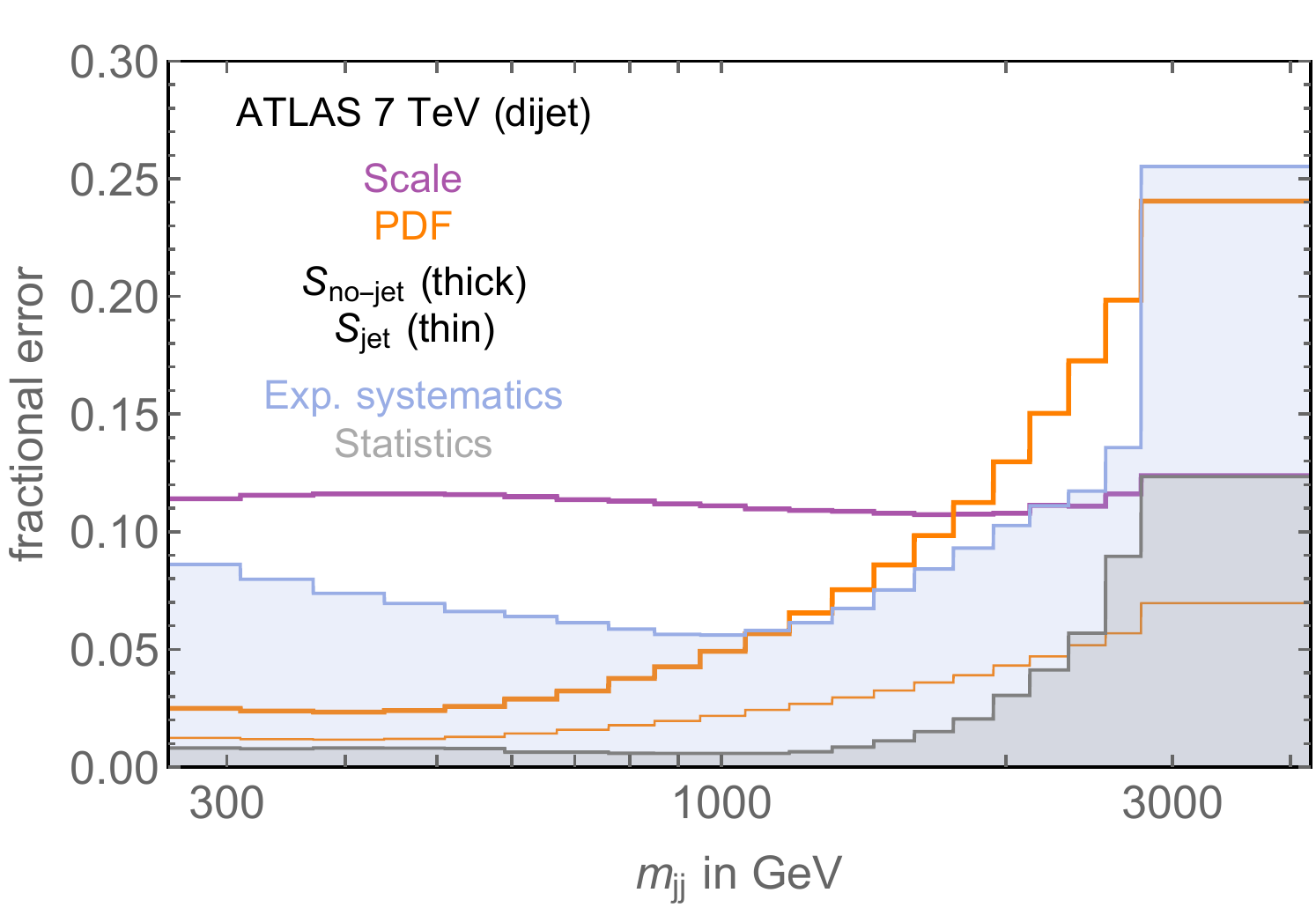}~~~~~\includegraphics[width=0.452\textwidth]{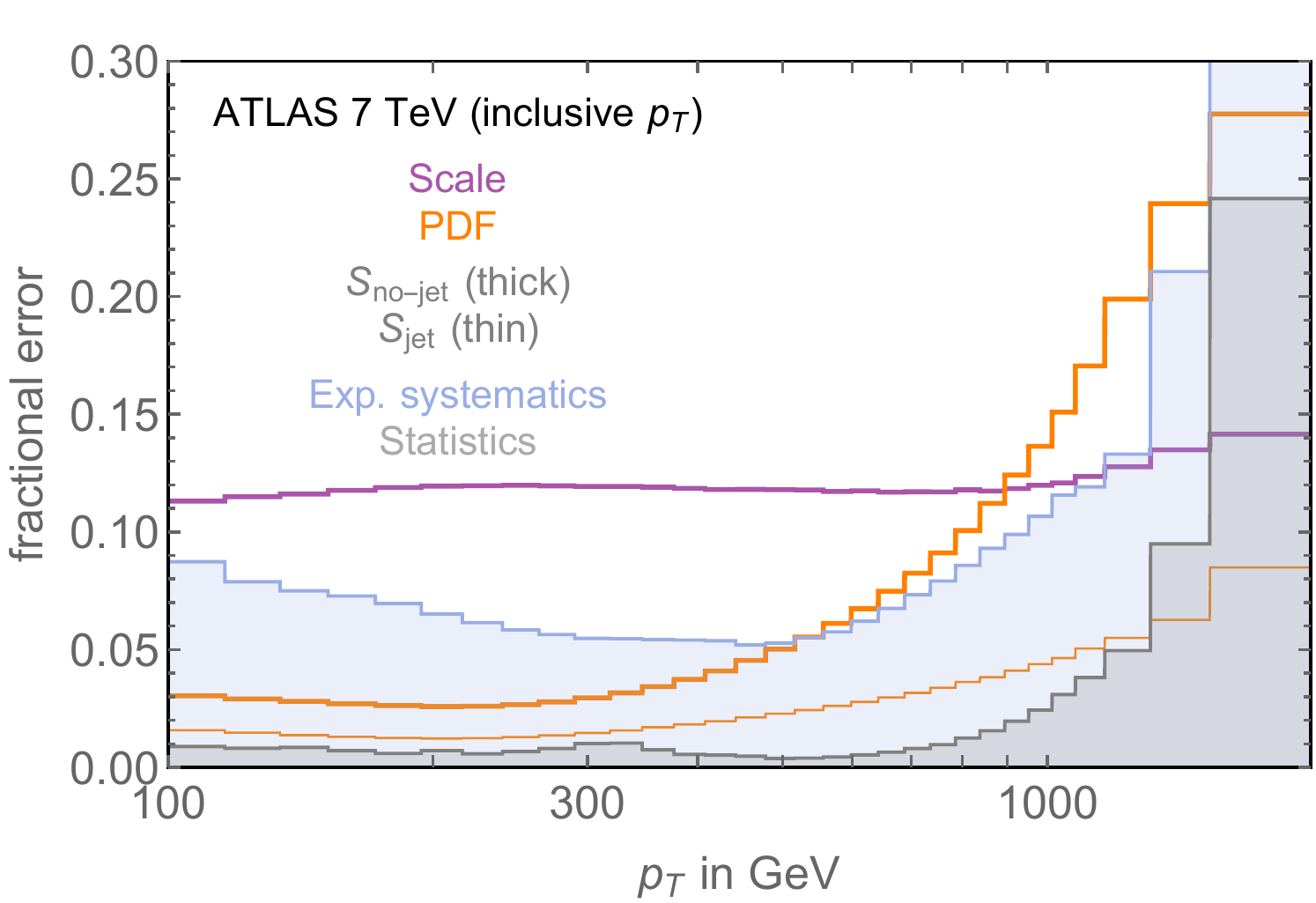}
\end{center}
\vspace{-.3cm}
\caption{ \footnotesize
Scale and PDF uncertainties for the central rapidity bin of the ATLAS dijet (left) and inclusive $p_T$ (right) analyses. Similar results holds for less central bins and for the analogue CMS analyses. The scale uncertainty is flat in energy and dominates at low invariant masses ($p_T$). The PDF uncertainty grows with the energy and dominates over the scale one for the \texttt{NNPDF30\_nlo\_as\_0118\_nojet} PDF set ($\mathcal S_{\textrm{no-jet}}$). The PDF uncertainty associated to the \texttt{NNPDF30\_nlo\_as\_0118} PDF set ($\mathcal S_{\textrm{jet}}$) is roughly one half of the one associated to $\mathcal S_{\textrm{no-jet}}$.
}
\label{errors}
\end{figure}
%%%%%%%%%%%%%
%%%%%%%%%%%%%
%%%%%%%%%%%%%
%%%%%%%%%%%%%

We consider four sources of theoretical uncertainty: scale variations, PDFs, $\alpha_s$ choice and showering and hadronization effects.
We evaluate the scale uncertainty on the cross section by varying the renormalization and factorization scale, $\mu_r$ and $\mu_f$. 
We consider seven choices of scales $(\mu_r,\mu_f)=(a,b)\times p_T^{(uB)}$, where $a,b=1/2,1,2$, $(a,b)\neq (1/2,2),(2,1/2)$, and $p_T^{(uB)}$ is the underlying-Born transverse momentum as defined for instance in Ref.~\cite{Alioli:2010xa}. The scale uncertainty in the $i$-th bin is taken to be $(\sigma_i^{\textrm{max}}-\sigma_i^{\textrm{min}})/2$, where $\sigma_i^{\textrm{max}}$ and $\sigma_i^{\textrm{min}}$ are the maximal and minimal value of the cross section that are achieved among the seven choices of renormalization and factorization scale. The scale uncertainty is fully correlated across all bins.

PDFs uncertainties and correlations across all energy and rapidity bins are evaluated using the standard NNPDF replicas prescription.\footnote{Each NNPDF set comes with a central value and a set of replicas. Given two bins $i$ and $j$ in which a cross section has to be calculated, the $ij$ element of the PDF covariance matrix is given by
\be
{\textrm{cov}}_{ij}=N_R^{-1}\sum_r (\sigma_i^{(r)}-\bar\sigma_i)(\sigma_j^{(r)}-\bar\sigma_j)
\ee
where $\bar\sigma_i$ and $ \sigma_i^{(r)}$ are the cross section in the $i$-th bin calculated with the central value PDF and the $r$-th replica respectively. $N_R=100$ is the number of replicas in a given NNPDF set.} 

The $\alpha_s$ uncertainty is evaluated using the PDF4LHC recommendation~\cite{Butterworth:2015oua}. Two additional PDF sets are provided employing two different values of the strong coupling constant, $\alpha_s^{+(-)}=0.1195(65)$. Similarly to the scale uncertainty, for any bin $i$ we take the uncertainty on the cross section associated with $\alpha_s$ to be given by $(\sigma_i^{\textrm{max}}-\sigma_i^{\textrm{min}})/2$ where now $\sigma_i^{\textrm{max}}$ and $\sigma_i^{\textrm{min}}$ are the maximal and minimal value of $\sigma_i$ achieved with the three different choices of $\alpha_s$. This source of uncertainty is completely correlated across all bins. $\alpha_s$ uncertainties are always small, of order few percents, and have a negligible effect on our fits. For this reason we do not include them in the following.

Finally, in order to estimate the uncertainty due to showering and hadronization we perform the showering and hadronization stages with two different shower Monte Carlo programs, {\tt{Pythia6}}~\cite{Sjostrand:2006za} and {\tt{Pythia8}}~\cite{Sjostrand:2014zea} and with three different tunes.  We associate an error to the result $\sigma_i$ obtained for each shower program and tune choice and we perform a convolution of the results to produce an error estimate using the same procedure we used for scale uncertainty. The shower and hadronization errors obtained in this way are included in the fit. For the observables under consideration this uncertainty is usually small, of order few percents.

In Fig.~\ref{errors} we display the fractional value of the leading theoretical uncertainties for the dijet and inclusive $p_T$ analyses performed by ATLAS, in the central rapidity bin. Similar results hold for the analogue analyses performed by CMS\@. The scale uncertainty is almost constant in energy, of order 10\% and dominates at low energies. For the PDF uncertainty, we show two possible PDF choices, both from NNPDF3.0. The first one, dubbed $\mathcal S_{\textrm{jet}}$, is the standard NLO set with $\alpha_s(m_Z)=0.118$ and set name \texttt{NNPDF30\_nlo\_as\_0118}. The second one, corresponding instead to set name \texttt{NNPDF30\_nlo\_as\_0118\_nojet} and dubbed $\mathcal S_{\textrm{no-jet}}$, does not include jet data, both from Tevatron~\cite{Abulencia:2007ez} and low luminosity Run-I LHC (ATLAS inclusive jet at 2.76 and 7 TeV~\cite{Aad:2013lpa}, CMS 7 TeV inclusive jet and dijet~\cite{Chatrchyan:2012bja} and $t\bar t$ production at ATLAS~\cite{ATLAS:2011xha,ATLAS:2012aa,TheATLAScollaboration:2013dja} and CMS~\cite{Chatrchyan:2012bra, Chatrchyan:2012ria,Chatrchyan:2013faa}.) For both sets, the PDF uncertainty is negligible at low energies, but can become the dominant source of theory uncertainty at high energy in the case of $\mathcal S_{\textrm{no-jet}}$. The fractional uncertainty associated with $\mathcal S_{\textrm{jet}}$ is roughly half the size of the one from $\mathcal S_{\textrm{no-jet}}$, throughout the energy range. 
The reason is the gluon PDF, which is strongly constrained by the jet data included in $\mathcal S_{\textrm{jet}}$. 

The large uncertainty associated with $\mathcal S_{\textrm{no-jet}}$ introduces an important limitation in trying to constrain the SM EFT by measuring the high energy tail of some kinematical distribution involving jets.
One might be tempted to only use $\mathcal S_{\textrm{jet}}$, given its smaller uncertainty, but there is a possible concern in doing so. If NP is present, it could contaminate the same jet data that are used in $\mathcal S_{\textrm{jet}}$ to constrain the PDFs. Using $\mathcal S_{\textrm{jet}}$ to constrain this same NP would then be circular and potentially result in an artificially strong bound.  However, we point out that contact operators in general, and Eq.~\ref{LZ} in particular, give their largest contributions in the central kinematical region, while PDF extraction is typically more sensitive to the forward region, where the cross section is larger. Given this caveat, we will proceed by using both sets $\mathcal S_{\textrm{jet}}$ and $\mathcal S_{\textrm{no-jet}}$ in the following, as comparitive benchmarks to estimate the sensitivity to the $\rmZ$ parameter.

\begin{figure}[H]
\vspace{-2.1cm}
\begin{center}
\includegraphics[width=0.95\textwidth]{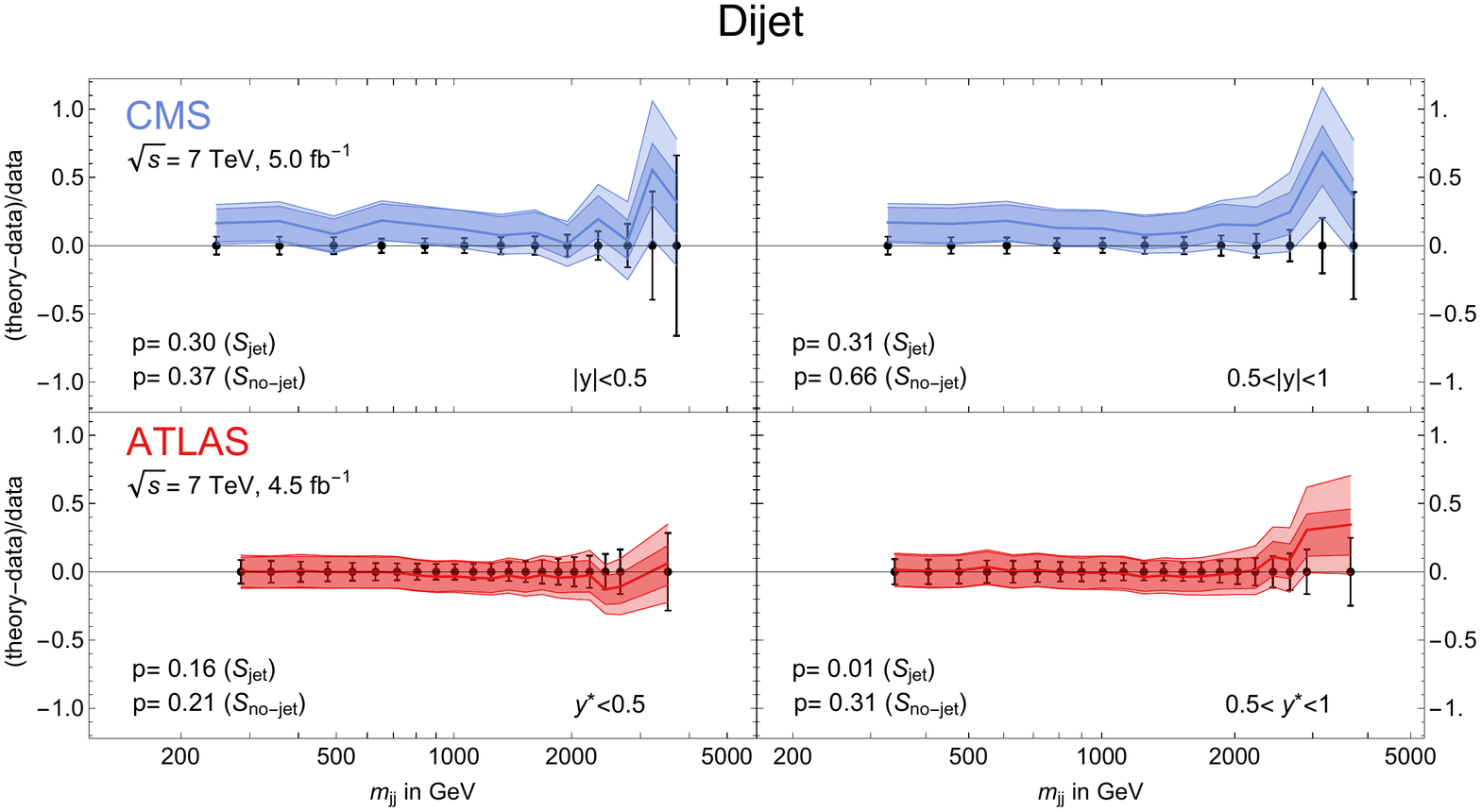}
\end{center}
\vspace{-4.1cm}
\begin{center}
\includegraphics[width=0.95\textwidth]{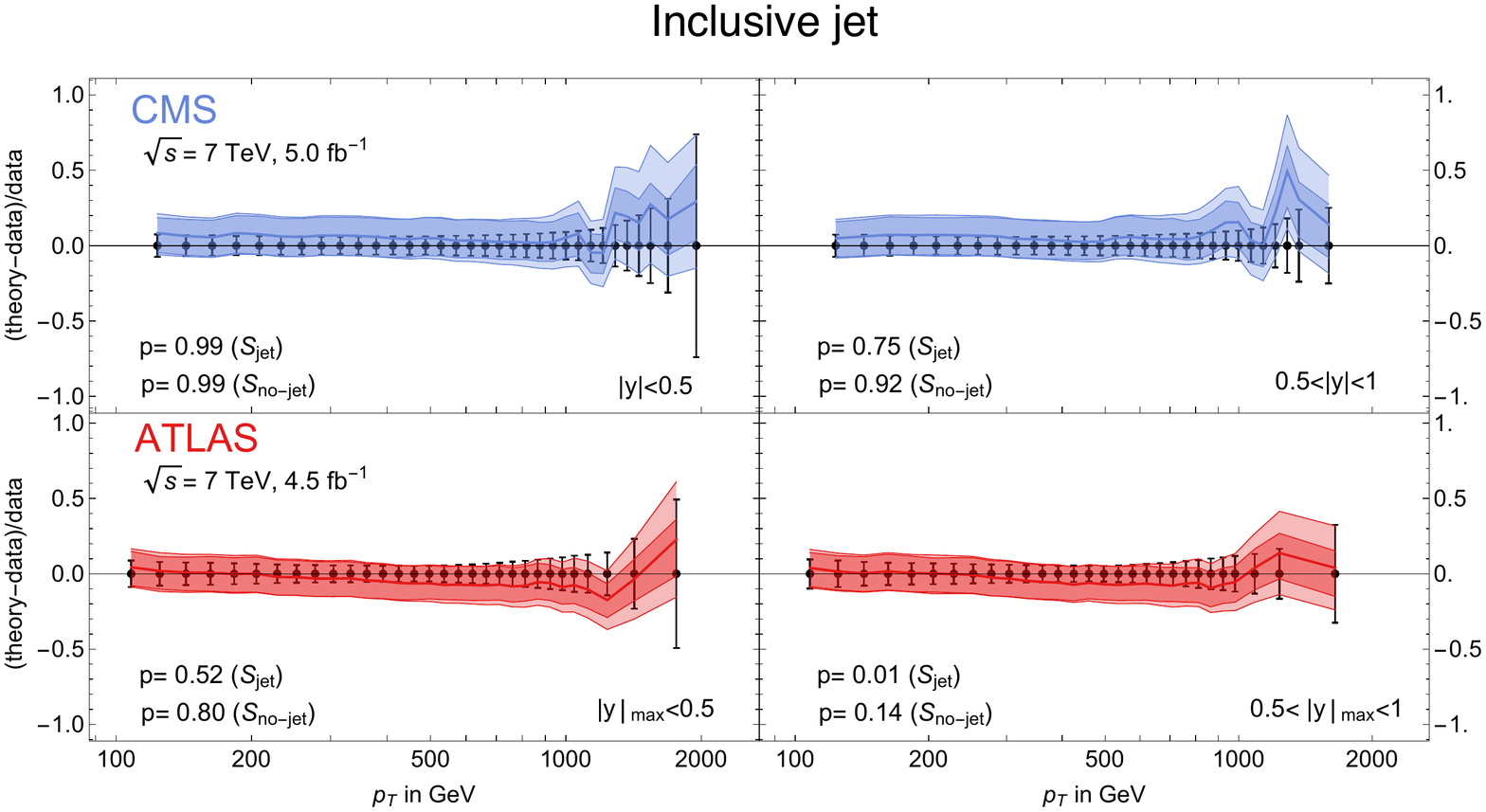}
\end{center}
\vspace{-1.5cm}
\caption{ \footnotesize
Comparison of our theory predictions with the experimental data. We show the results for the two most central rapidity bins of each analysis. The black error bars represent the experimental uncertainties. The shaded regions represent the sum in quadrature of all theory uncertainties. The lighter shaded regions correspond to $\mathcal S_{\textrm{nojet}}$ while the darker one to $\mathcal S_{\textrm{jet}}$. The theory prediction is the thick colored line. For each bin we show the p-value that results from Eq.~\ref{chiSM}.
}
\label{fitSM}
\end{figure}

In Fig.~\ref{fitSM}, we compare our calculations with the experimental results for both dijet and inclusive $p_T$, for both ATLAS and CMS, and for the two most central rapidity bins. While the black error bars represent the fractional size of the $1\sigma$ experimental uncertainties, the shaded region displays the fractional size of the theory uncertainty, calculated as the sum in quadrature of all the effects described above. The two theory uncertainty bands correspond to the two choices of PDF sets, the wider one being associated to $\mathcal S_{\textrm{no-jet}}$ and the smaller one to $\mathcal S_{\textrm{jet}}$.

In order to get a sense of the quality of the fit we build a $\chi^2$ statistic
\be\label{chiSM}
\chi^2=\sum_{i,j}(\sigma^{\textrm{th}}_i-\sigma^{\textrm{exp}}_i)(\Sigma^{-1})_{ij}(\sigma^{\textrm{th}}_j-\sigma^{\textrm{exp}}_j)
\ee
where $\sigma^{\textrm{th}}_i$ and $\sigma^{\textrm{exp}}_i$ are the theory and experimental cross section in the $i$-th bin, respectively, and $\Sigma$ is the uncertainty covariance matrix constructed as
\be\label{cov1}
\Sigma=\Sigma^{\textrm{exp}}+\Sigma^{\textrm{th}}.%+\Sigma^{\textrm{PDF}}+\Sigma^{\alpha_s}+\Sigma^{\textrm{had}}.
\ee
$\Sigma^{\textrm{exp}}$ is the covariance matrix provided by the experimental collaboration and $\Sigma^{\textrm{th}}$ is the theory covariance matrix including scale, PDF, $\alpha_s$, and hadronization uncertainties.  We notice that with two exceptions (the $0.5\leq y^*<1$ bin in the ATLAS dijet and inclusive jet measurements when fitted with $\mathcal S_{\textrm{jet}}$) all the p-values show good consistency between theory and data.

%%%%%%%
%%%%%%%%%%%%%
%%%%%%%%%%%%%
%%%%%%%%%%%%%
%%%%%%%%%%%%%
%%%%%%%%%%%%%
%%%%%%%%%%%%%
%%%%%%%

\begin{figure}[t]
\begin{center}
~~\includegraphics[width=1.0\textwidth]{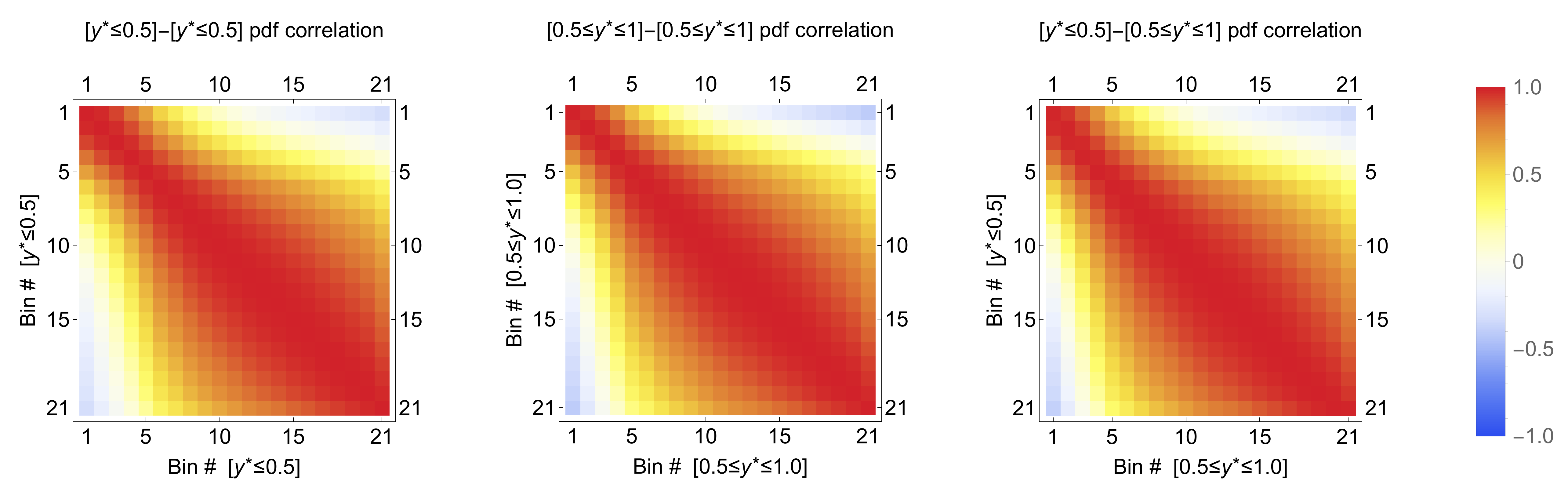}
\end{center}
\vspace{-.3cm}
\caption{ \footnotesize
PDF correlation (from the $\mathcal S_{\textrm{no-jet}}$ set) for the 7 TeV ATLAS dijet analysis. We show the correlation among different invariant mass bins in the first two rapidity bins (first and second panel) and the cross correlation between the first and second rapidity bin (third panel). Crucially, different rapidity bins with similar invariant mass have highly correlated PDF uncertainties.
}
\label{PDFcorr}
\end{figure}

The results of the double-differential fit are presented in Appendix~\ref{app:ddf}. We briefly discuss them here to point out their main features. While the fits to each individual rapidity bin have a reasonable p-value, the p-values worsen when one tries to fit multiple rapidity bins. This is especially the case for ATLAS inclusive $p_T$ and dijet, where including more than two rapidity bins leads to a vanishingly small p-value. The possibility of this behavior is due to the existence of correlated uncertainties. Including only the central rapidity bin of one analysis, the main source of uncorrelated uncertainty is from the PDF\@.  When two or more rapidity bins are used, one has to take into account the cross correlation of the associated PDF uncertainty. This cross correlation is shown in Fig.~\ref{PDFcorr}.

As expected, in a given rapidity bin the PDF uncertainties are highly correlated for nearby energy bins. What may be surprising is that nearby energy bins are also highly correlated across different rapidity bins. The reason for this is that for a given center of mass energy, $\sqrt{s}$, and a given $m_{jj}$ bin,\footnote{A similar argument applies for the $p_T$ distribution.} approximately a single value of $x_1 x_2=m^2_{jj}/s$ is sampled by the PDFs (where $x_1$ and $x_2$ are the two parton momentum fractions).
When fitting a double differential distribution, part of the uncertainty associated to the PDFs will drop because of this correlation, and this can lead to a worsening of the fit.
Clearly such a degradation is only expected if some of the uncertainties (either theoretical or experimental) have been underestimated. We are not able to pin point with certainty the source of the problem, which seems to mainly affect the ATLAS results. We note that Ref.~\cite{Malaescu} makes a similar observation, when fitting the inclusive jet $p_T$ double differential distribution at 8\,TeV.

We  stress, however,  that we find good agreement between data and our calculation in the central region of each search. This agreement persists if we  fit the ATLAS dijet and inclusive jet data using a fixed order NLO calculation from {\tt{NLOJet++}}~\cite{Nagy:2003tz}. The details of this additional check can be found in Appendix~\ref{app:ddf}.

%%%%%%%%%%%%%
%%%%%%%%%%%%%
%%%%%%%%%%%%%
%%%%%%%%%%%%%
%%%%%%%%%%%%%
%%%%%%%%%%%%%
%%%%%%%%%%%%%
%%%%%%%%%%%%%
We can now proceed to constrain the operator in Eq.~\ref{LZ}. For a given experimental measurement, the cross section in a certain energy bin is quadratic in the $\rmZ$ parameter
\be\label{sigmaZ}
\sigma^{\textrm{th}}_i(\rmZ)=\sigma^{SM}_i+\sigma^{(1)}_i\rmZ+\sigma^{(2)}_i\rmZ^2.
\ee
$\sigma^{SM}_i$ is the value of the SM cross section in the $i$-th bin. For each analysis, we use the leading order formulas in Table~\ref{tab:M2Z} to calculate the values of $\sigma_i^{(1)}$ and $\sigma_i^{(2)}$, by integrating the partonic cross sections over the phase space defined by the experimental cuts for each analyses. The PDF integration is performed in \textsc{Mathematica} using the available PDF grids from the LHAPDF set~\cite{Buckley:2014ana}. The inclusion of the NLO correction to the BSM calculation can result in a reduction of the BSM cross section by 10$-$20\% at large $m_{jj}/p_T$~\cite{Gao:2011ha}.

This will result in a similar weakening of the bound on $\rmZ$. We notice that these BSM NLO effects will only represent a small multiplicative correction of order 10\% to the NP scale which is constrained by the searches. For this reason we consider their inclusion of lower priority with respect to the evaluation of SM NLO effects and their associated uncertainties.

%%%%%%%%%%%%%%%%%%%%%%%%%%%%%%%%%%%%%%%%%%%%%%%%%%%%%%%%%%%%%%%%%%%%%%%%%%
%%%%%%%%%%%%%%%%%%%%%%%%%%%%%%%%%%%%%%%%%%%%%%%%%%%%%%%%%%%%%%%%%%%%%%%%%%
\begin{table}[t]
\caption*{95\%~CL bounds on $\rmZ\times 10^4$}
\begin{center}\vspace{-0.5cm}
{\small
\begin{tabular}{c|c|c|c} 
Analysis & $\mathcal S_{\textrm{no-jet}}$ - 1bin & $\mathcal S_{\textrm{no-jet}}$ - 2bins& $\mathcal S_{\textrm{jet}}$ - 1bin\\ \hline \hline
ATLAS dijet &[-19.8,+3.9] & $^*$[+4.1,+9.3] & [-5.4,+4.3] \\ 
ATLAS inclusive jet&[-18.7,+8.7] & [-6.5,+1.3]&[-3.2,+8.0] \\
CMS dijet &[-18.3,+6.0] &[-2.3,+5.5] & [-5.5,+2.8]\\
CMS inclusive jet & [-18.9,+3.1]& $^*$[-8.6,-0.4]& [-8.0,+1.9] \\
\end{tabular}
} 
\vspace{0.3cm}
\caption{\label{results1}\footnotesize 95\% CL constraints on the $\rmZ$ parameter. Different rows correspond to the four experimental searches of Table~\ref{tab:searches}, while different columns correspond to different PDF sets or different numbers of rapidity bins included in the fit. The entries marked with `*'  are those for which the fit excludes the SM ($\rmZ=0$) at 95\% CL\@.}
\end{center}
\end{table}
%%%%%%%%%%%%%%%%%%%%%%%%%%%%%%%%%%%%%%%%%%%%%%%%%%%%%%%%%%%%%%%%%%%%%%%%%%
%%%%%%%%%%%%%%%%%%%%%%%%%%%%%%%%%%%%%%%%%%%%%%%%%%%%%%%%%%%%%%%%%%%%%%%%%%

In order to constrain $\rmZ$ we construct a likelihood function
\be\label{llh1}
-2\log \mathcal{L}(\rmZ)=\sum_{i,j}(\sigma^{\textrm{th}}_i(\rmZ)-\sigma^{\textrm{exp}}_i)(\Sigma^{-1})_{ij}(\sigma^{\textrm{th}}_j(\rmZ)-\sigma^{\textrm{exp}}_j)
\ee
where $\Sigma$ is the same covariance matrix of Eq.~\ref{cov1}. Using the normalized likelihood function $\hat{\mathcal L}(Z)$ as the posteriori probability distribution for $\rmZ$ (assuming a flat prior on $\rmZ$) we calculate Confidence Level (CL) intervals for $\rmZ$ as the iso-contours of $\mathcal L$ containing a given probability $p$.

We validate our procedure by reproducing the contact operator bounds set by ATLAS~\cite{Aad:2013tea} and CMS using the same analyses~\cite{Chatrchyan:2013muj}. The results of our validation are shown in Appendix~\ref{app:validation} and they are consistent with the results of the experimental collaborations.

Our ability to constrain $\rmZ$ relies on the fact that the ratios $\sigma^{(1)}_i/\sigma^{SM}_i$ and $\sigma^{(2)}_i/\sigma^{SM}_i$ grow as the appropriate energy variable, either the invariant mass or the transverse momentum, is increased. In particular the effect of $\rmZ$ is such that positive values of $\rmZ$ correspond to a positive interference with the SM ($\sigma^{(1)}_i>0$), while negative values correspond to destructive interference ($\sigma^{(1)}_i<0$).

We display the results of the fit in Table~\ref{results1}. We present them in three different ways corresponding to different choices of the PDF set and the number of rapidity bins included in the fit.
Using only the most central bin of each analysis and $\mathcal S_{\textrm{no-jet}}$, we find a very similar constraint on $\rmZ$. We notice in particular that the 95\% CL interval is asymmetric, with the limit being weaker for $\rmZ<0$ where there is negative interference with the SM\@. This is due to the fact that the large systematic uncertainties, in particular the PDF ones, make the likelihood non-Gaussian, and the limit non-symmetric.
We also present the same results using either $\mathcal S_{\textrm{no-jet}}$ and two rapidity bins, or $\mathcal S_{\textrm{jet}}$ and a single rapidity bin. 

As explained above, the fits combining two rapidity bins have, in general,  lower p-values at the $\rmZ=0$ point. This in particular implies that the ATLAS dijet and CMS inclusive-jet fit exclude the SM at the 95\% CL\@. We notice however that for the other searches the constraint improves significantly with respect to $\mathcal S_{\textrm{no-jet}}$/1\,bin, and the reason for this is the large correlation of the PDF uncertainties across different rapidity bins. We do not show the results of the fit by adding more rapidity bins, because 
%for those cases for which the SM p-value is reasonably good
we find the addition of additional bins, beyond the two most central, to have a negligible impact on the constraint on $\rmZ$.

%%%%%%%%%%%%%
%%%%%%%%%%%%%
%%%%%%%%%%%%%
%%%%%%%%%%%%%
\begin{figure}[t]
\begin{center}
~~\includegraphics[width=0.472\textwidth]{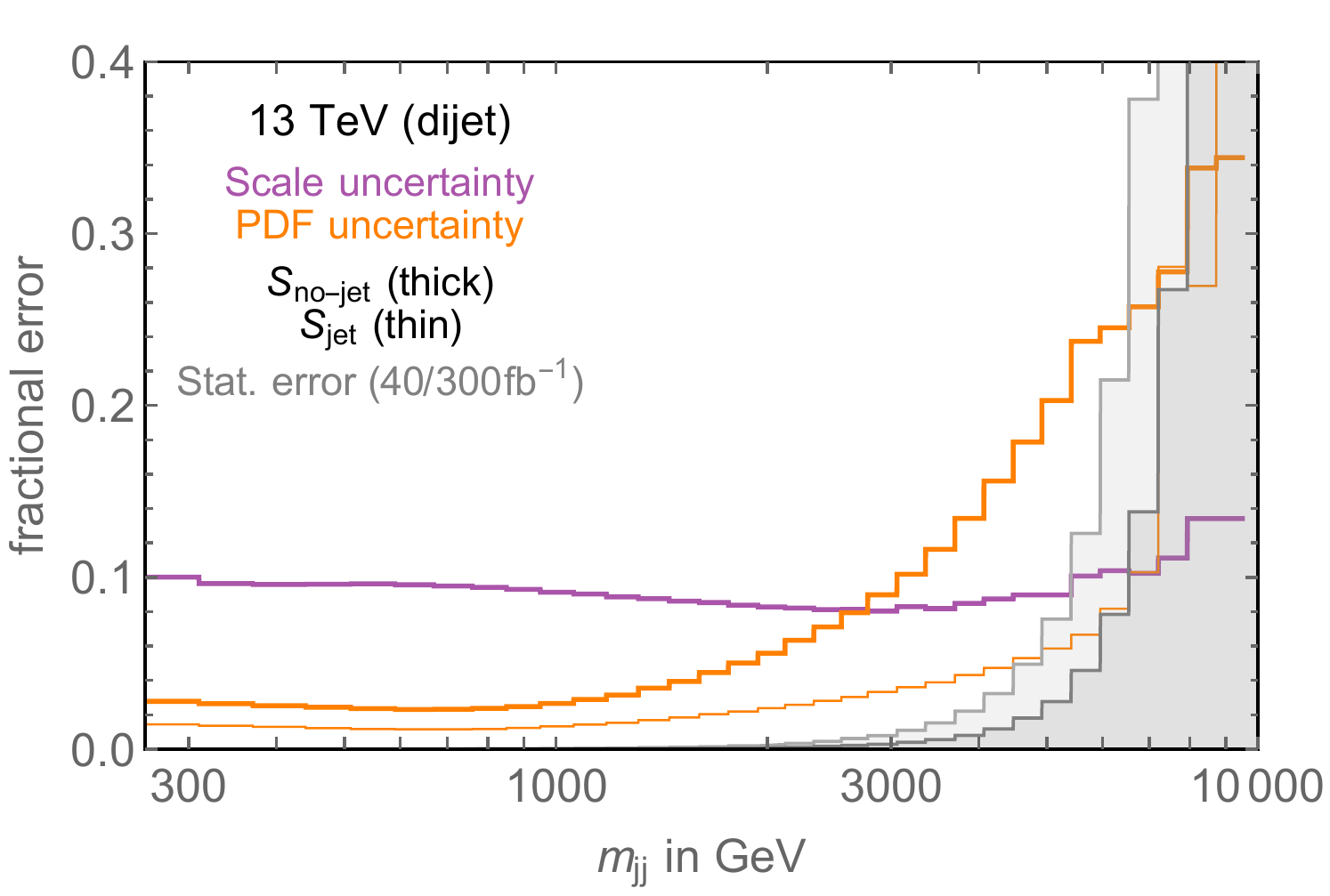}~~~~~\includegraphics[width=0.452\textwidth]{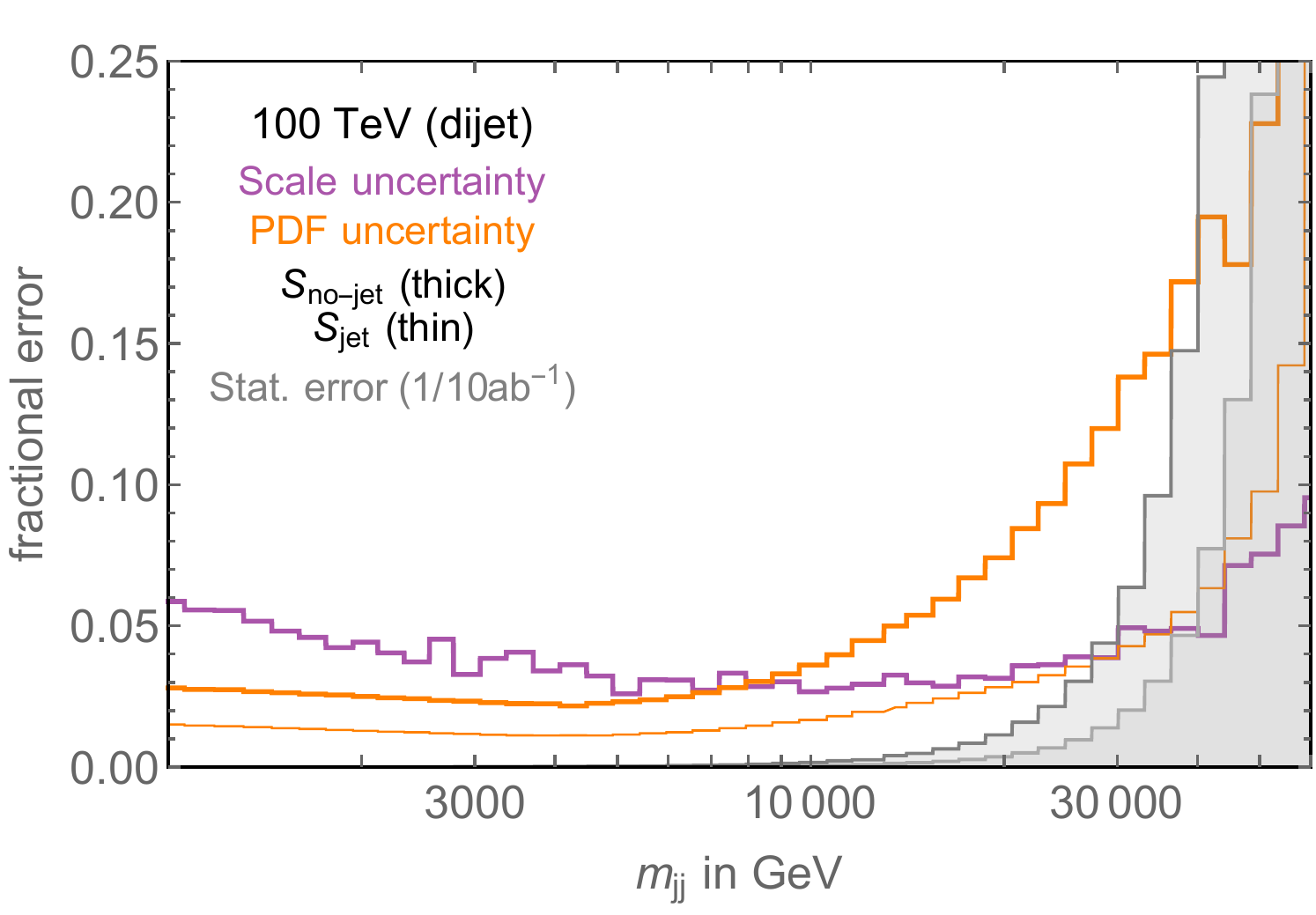}
\end{center}
\vspace{-.3cm}
\caption{ \footnotesize
Scale and PDF uncertainties for the central rapidity bin of the 13\,TeV (left) and 100\,TeV (right) dijet analyses. Similar results hold for for the inclusive jet $p_T$ analyses identifying $p_T\sim m_{jj}/2$. We also show the statistical uncertainty for two values of the integrated luminosity.
}
\label{errorsproj}
\end{figure}
%%%%%%%%%%%%%
%%%%%%%%%%%%%
%%%%%%%%%%%%%
%%%%%%%%%%%%%

For the $\mathcal S_{\textrm{jet}}$/1\,bin fit, the size of the 95\% CL interval is again smaller than the one of the $\mathcal S_{\textrm{no-jet}}$/1\,bin fit, and comparable to that of the $\mathcal S_{\textrm{no-jet}}$/2\,bin one. Again, including additional rapidity bins do not significantly impact the bound.

We conclude this section by noting that, already with half of the full LHC center of mass energy and a small fraction of the final available luminosity, jet physics is able to place powerful constraints on modifications of the QCD sector, such as those represented by Eq.~\ref{LZ}. The constraint on the $\rmZ$ parameter is at the level of $5\times10^{-4}$ level, corresponding to a mass scale $M\equiv m_W/\sqrt{Z}\approx 3.5$\,TeV\@.

%%%%%%%%%%%%%%%%%%%%%%%%%%%%%%%%%%%%%%%%%%%%%%%%%%%%%%%%%%%%%%%%%%%%%%%%%%
\section{Projected reach}
\label{sec:bounds2}
%%%%%%%%%%%%%%%%%%%%%%%%%%%%%%%%%%%%%%%%%%%%%%%%%%%%%%%%%%%%%%%%%%%%%%%%%

In this section, we project the reach on $\rmZ$ to values of the center of mass energies and luminosities for which either the full statistical information regarding the measurements, or the measurements themselves, are not yet available. In particular, we show the projected reach for a possible future 100\,TeV $pp$ collider.

As for the 7\,TeV dataset, we consider two different kinematical distributions: dijet invariant mass and inclusive jet $p_T$. At 8 and 13\,TeV, the analyses are defined using the same kinematic cuts as the ATLAS ones listed in Table~\ref{tab:searches}.  At 100\,TeV, the $\eta$ cut for the dijet analysis is raised to $|\eta| < 5$.  We proceed to calculate a $\rmZ$ dependent prediction for all the distributions, at the different energies, in the same way we did in the previous section for the 7\,TeV analyses. In Fig.~\ref{errorsproj}, we show the breakdown of the theory uncertainties for the central rapidity bin of the dijet invariant mass distribution, at both 13 and 100\,TeV\@. The behavior of the various uncertainties is indeed similar to the 7\,TeV analysis, as shown in Fig.~\ref{errors}.

%%%%%%%%%%%%%%%%%%%%%%%%%%%%%%%%%%%%%%%%%%%%%%%%%%%%%%%%%%%%%%%%%%%%%%%%%%
%%%%%%%%%%%%%%%%%%%%%%%%%%%%%%%%%%%%%%%%%%%%%%%%%%%%%%%%%%%%%%%%%%%%%%%%%%
\begin{table}[t]
\caption*{95\%~CL bounds on $\rmZ\times 10^4$ for $\sqrt s=8$\,TeV}
\begin{center}\vspace{-0.5cm}
{\small
\begin{tabular}{c|c|c|c} 
Analysis & $\mathcal S_{\textrm{no-jet}}$ - 1bin & $\mathcal S_{\textrm{no-jet}}$ - 2bins& $\mathcal S_{\textrm{jet}}$ - 1bin\\ \hline \hline
dijet &[-9.4,+4.9] & [-2.6,+2.1]  & [-2.1,+1.8] \\ 
inclusive jet&[-13.8,+4.2] & [-2.5,+2.3]&[-2.7,+2.1] \\
\end{tabular}
} 
\vspace{0.3cm}
\caption{\label{results8}\footnotesize Projected 95\% CL constraints on the $\rmZ$ parameter for the 8\,TeV LHC with 20\,fb$^{-1}$ of integrated luminosity. The two rows correspond to the limit extracted from either the dijets or the inclusive jet analysis, respectively,
while the three different columns correspond to different PDF sets or different number of rapidity bins included in the fit.}
\end{center}
\end{table}
%%%%%%%%%%%%%%%%%%%%%%%%%%%%%%%%%%%%%%%%%%%%%%%%%%%%%%%%%%%%%%%%%%%%%%%%%%
%%%%%%%%%%%%%%%%%%%%%%%%%%%%%%%%%%%%%%%%%%%%%%%%%%%%%%%%%%%%%%%%%%%%%%%%%%

In order to estimate the reach on $\rmZ$ we consider the following likelihood function
\be\label{llh2}
-2\log \mathcal{L}(\rmZ)=\sum_{i,j}(\sigma^{\textrm{th}}_i(\rmZ)-\sigma^{SM}_i)(\Sigma^{-1})_{ij}(\sigma^{\textrm{th}}_j(\rmZ)-\sigma^{SM}_j).
\ee
The covariance matrix, $\Sigma$, is given by
\be
\Sigma=\Sigma^{\textrm{stat}}+\Sigma^{\textrm{exp}}+\Sigma^{\textrm{th}}.
\ee
Like in the previous section, $\Sigma^{\textrm{th}}$ is the theory covariance matrix including scale, PDF, $\alpha_s$, and hadronization uncertainties. As an estimate of the experimental uncertainties, we assume $\Sigma^{\textrm{exp}}$ to be the sum of two components, one completely uncorrelated ($\delta_U$) and one completely correlated ($\delta_C$)
\be\label{experrmock}
(\Sigma^{\textrm{exp}})_{ij}=(\delta^2_C+\delta_{ij}\delta^2_U)\sigma^{\textrm{th}}_i \sigma^{\textrm{th}}_j \, .
\ee
At 8 and 13\,TeV, we take $(\delta_U,\delta_C)=(2\%,5\%)$, while at 100\,TeV, $(\delta_U,\delta_C)=(5\%,10\%)$. We validated this choice by redoing our 7 TeV, replacing the true experimental uncertainties with the fixed values $(\delta_U,\delta_C)=(2\%,5\%)$.  We verify that the bounds of Table~\ref{results1} are approximately replicated with this simplified treatment of experimental uncertainties. We opted for more conservative uncertainties for our 100\,TeV projections. The statistical part of the covariance matrix is defined by
\be
\Sigma^{\textrm{stat}}_{ij}=\frac{\sigma^{\textrm{th}}_i(\rmZ)}{{L}}\delta_{ij}
\ee
where $L$ is the integrated luminosity.

%%%%%%%%%%%%%
%%%%%%%%%%%%%
%%%%%%%%%%%%%
%%%%%%%%%%%%%
\begin{table}[t]
\caption*{95\%~CL bounds on $\rmZ\times 10^4$ for $\sqrt s=13,100$\,TeV}
\begin{center}\vspace{-0.5cm}
{\small
\begin{tabular}{c|c|c|c|c} 
Analysis & $\sqrt{s}$ $-$ Luminosity & $\mathcal S_{\textrm{no-jet}}$ - 1bin & $\mathcal S_{\textrm{no-jet}}$ - 2bins& $\mathcal S_{\textrm{jet}}$ - 1bin\\ \hline \hline
\multirow{ 4}{*}{dijet } & $13 \,\text{TeV}$  $-$ $40\,\text{fb}^{-1}$ &[-3.3,+1.7] & [-1.0,+0.9]  & [-0.8,+0.7] \\ 
&$13 \,\text{TeV}$   $-$ $0.3 \,\text{ab}^{-1}$  &[-3.1,+1.4] & [-0.7,+0.6]&[-0.6,+0.5] \\
&$13 \,\text{TeV}$ $-$ $3 \,\text{ab}^{-1}$ &[-2.8+1.2] & [-0.5,+0.4]&[-0.5,+0.5] \\
&$100 \,\text{TeV}$  $-$ $10 \,\text{ab}^{-1}$ &[-4.5,+2.5]$\times 10^{-2}$ & [-2.4,+1.7] $\times 10^{-2}$&[-1.4,+1.2] $\times 10^{-2}$ \\ \hline
\multirow{ 4}{*}{inclusive jet } & $13 \,\text{TeV}$  $-$ $40 \,\text{fb}^{-1}$ &[-5.0,+1.5] & [-1.0,+0.9]  & [-1.0,+0.8] \\ 
&$13 \,\text{TeV}$  $-$ $0.3 \,\text{ab}^{-1}$ &[-4.2,+1.1] & [-0.7,+0.6]&[-0.7,+0.6] \\
&$13 \,\text{TeV}$  $-$ $3 \,\text{ab}^{-1}$ &[-3.5,+0.9] & [-0.5,+0.5]&[-0.6,+0.5] \\
&$100 \,\text{TeV}$  $-$ $10 \,\text{ab}^{-1}$ &[-10.7,+2.6]$\times 10^{-2}$  & [-1.6,+1.4]$\times 10^{-2}$ &[-1.9,+1.5]$\times 10^{-2}$  \\
\end{tabular}
} 
\vspace{0.3cm}
\caption{\label{results13}\footnotesize Projected 95\% CL constraints on the $\rmZ$ parameter for the LHC at 13\,TeV and a future circular $pp$ collider with $\sqrt s=100$\,TeV\@.}.
\end{center}
\end{table}
%%%%%%%%%%%%%
%%%%%%%%%%%%%
%%%%%%%%%%%%%
%%%%%%%%%%%%%

In Table~\ref{results8} we show the 8\,TeV projections.
For the inclusive jet analysis, we use the binning and cuts of the very recent ATLAS analysis~\cite{Malaescu}, for which the statistical information regarding the correlation of experimental errors is not yet publicly available. For the dijet one, we instead adopt the same binning and cuts that we used at 7\,TeV\@. The projections show a similar constraining power of the dijet and inclusive jet analyses. Furthermore, they show once again the crucial role of the PDF uncertainty in limiting the reach. This is evident from the different size of the 95\% CL interval extracted from $\mathcal S_{\textrm{no-jet}}$/1\,bin and $\mathcal S_{\textrm{no-jet}}$/2\,bin or $\mathcal S_{\textrm{jet}}/$1\,bin, the latter having comparable size $\rmZ\approx 2\times 10^{-4}$.

A similar trend emerges for our projections at 13 and 100\,TeV\@. The dijet and the inclusive jet search both have similar constraining power on $\rmZ$ at fixed center of mass energy and luminosity. We notice again that CLs extracted from the $\mathcal S_{\textrm{no-jet}}$/2\,bin and $\mathcal S_{\textrm{jet}}/$1\,bin fits are similar in size and stronger by roughly a factor of 2 than those extracted from the $\mathcal S_{\textrm{no-jet}}$/1\,bin fit.
One important message from our results is the role of the luminosity. At 13\,TeV the reach is of order $\rmZ\approx10^{-4}$ with 40\,fb$^{-1}$, and only improves by roughly 50\% with the full luminosity of HL-LHC, corresponding to a NP scale of order $M\equiv m_W/\sqrt\rmZ\approx 8$\,TeV . In particular we find almost no improvement in going from 300\,fb$^{-1}$ to 3\,ab$^{-1}$. This is attributed to the large size of the theory systematic uncertainties.

%%%%%%%%%%%%%
%%%%%%%%%%%%%
%%%%%%%%%%%%%
%%%%%%%%%%%%%
\begin{table}[t]
\caption*{95\%~CL bounds on $\rmZ\times 10^4$ for $\sqrt s$ combinations}
\begin{center}\vspace{-0.5cm}
{\small
\begin{tabular}{c|c|c} 
Analysis & $8+13$\,TeV  & $8+13+100$\,TeV\\ \hline \hline
dijet  & [-0.7,+0.6]  & [-1.0+0.9]$\times 10^{-2}$   \\ 
inclusive jet  & [-0.9,+0.7]  & [-1.2,+1.1] $\times 10^{-2}$ 
\end{tabular}
} 
\vspace{0.3cm}
\caption{\label{resultscombo}\footnotesize Projected 95\% CL constraints on $\rmZ$ obtained by combining the individual 8, 13, and 100\,TeV fits, including PDF correlations across different center of mass energies. In all cases, we use $\mathcal S_{\textrm{no-jet}}$ and include only the most central rapidity bin.}
\end{center}
\end{table}
%%%%%%%%%%%%%
%%%%%%%%%%%%%
%%%%%%%%%%%%%
%%%%%%%%%%%%%

Increasing the energy by roughly a factor of 8, going from LHC to a future 100\,TeV $pp$ collider, results on the other hand in a very strong sharpening of the bounds, by roughly a factor of 30, corresponding to $\rmZ\approx 1.5\times 10^{-6}$ and $M\gtrsim 60$\,TeV\@.

The similarity, observed throughout our study, between the results of the $\mathcal S_{\textrm{no-jet}}$/2\,bin fits and the $\mathcal S_{\textrm{jet}}/$1\,bin ones, implies that the double differential distribution of both dijets and inclusive jets is effectively able to constrain the PDFs, avoiding a possible contamination from NP\@. As we  discussed in the previous section, this happens because measurements of the cross section in the same invariant mass (or transverse momentum) bin, performed at different rapidities, provide two independent measurements of the same value of $x_1 x_2$.

\begin{figure}[t]
\begin{center}
~~\includegraphics[width=0.9\textwidth]{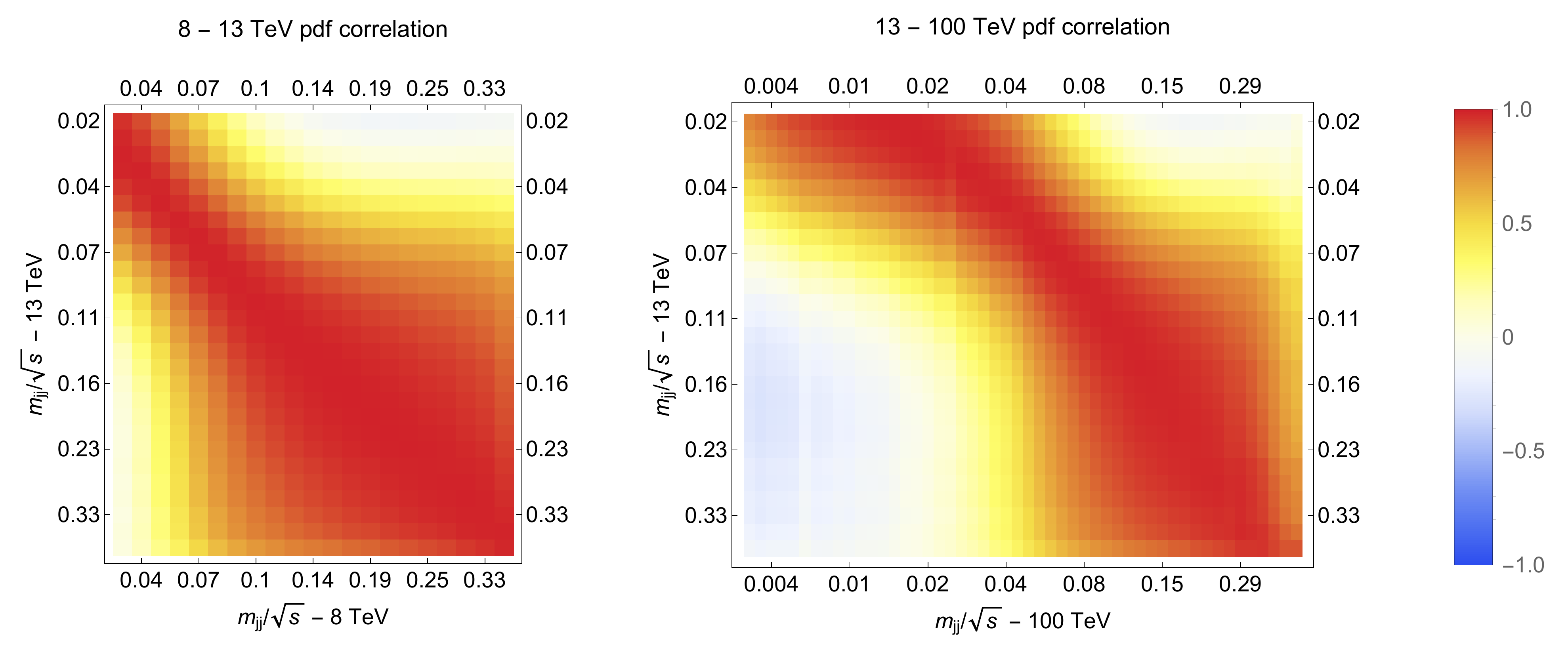}
\end{center}
\vspace{-.3cm}
\caption{ \footnotesize PDF correlation extracted from the $\mathcal S_{\textrm{no-jet}}$ set for the dijet analysis. We show how the PDF uncertainty is correlated across different invariant mass bins and collider energies (8\,-\,13\,TeV on the left and 13\,-100\,TeV on the right). The correlation is maximal for bins with similar values of $m_{jj}/\sqrt s$.
}
\label{PDFcorrcombo}
\end{figure}

We will now point out an alternative way to reduce the large PDF uncertainty associated to $\mathcal S_{\textrm{no-jet}}$: combining measurements performed at different collider center of mass energies. The reason for this can be exemplified by noticing that the dijet cross section at 8\,TeV, for $m_{jj}=2.5$\,TeV, constrain the same value of $x_1 x_2\equiv m^2_{jj}/s$ that is  constrained by a measurement performed at 13\,TeV for $m_{jj}=4$\,TeV\@. The existence of a large correlation of PDF uncertainties across bins with the same value of $m_{jj}/\sqrt s$ is shown in Fig.~\ref{PDFcorrcombo}.

We can thus use this observation and combine 8, 13, and 100\,TeV projections by carefully including their correlated PDF uncertainties.
We use $\mathcal S_{\textrm{no-jet}}$ and fit only the most central bin at each energy. The scale uncertainty is taken to be completely correlated across all bins included in the fit and the uncertainties in Eq.~\ref{experrmock} are included assuming no correlation across different center of mass energies.
The results of the $8+13$\,TeV and $8+13+100$\,TeV are shown in Table~\ref{resultscombo}, and they support our reasoning. The $8+13$\,TeV combination is as constraining as the $\mathcal S_{\textrm{no-jet}}$/2\,bin and $\mathcal S_{\textrm{jet}}/$1\,bin ones, while the $8+13+100$\,TeV delivers a slightly stronger bound.

%%%%%%%%%%%%%
%%%%%%%%%%%%%
%%%%%%%%%%%%%
%%%%%%%%%%%%%
%%%%%%%%%%%%%
%%%%%%%%%%%%%
%%%%%%%%%%%%%
%%%%%%%%%%%%%

\section{Validity of the effective field theory} \label{sec:eft}
One possible concern in using high energy processes to probe the SM EFT, is the fact that, like any EFT, its validity will break down at some high energy scale. In the case of the operator in Eq.~\ref{LZ}, this comes about because of its nature as a higher derivative correction to the gluon kinetic term.
The gluon propagator can be written down, at all orders in $\rmZ$, giving
\be\label{gluonPZfull}
P_G^T(q^2)=\frac{1}{q^2+\frac{\rmZ}{m_W^2}q^4}\,.
\ee
Expanding the above expression for small $Z$, one obtains Eq.~\ref{gluonPZ}. At an energy $E\sim m_W/\sqrt{Z}$, higher order corrections, for instance from operators of dimension 8 and above, are expected to become relevant, invalidating the effective description in term of Eq.~\ref{LZ}.\footnote{For $\rmZ<0$, Eq.~\ref{gluonPZfull} implies the existence of a ghost with mass $m_W/\sqrt{\rmZ}$. However, no general positivity argument exists to imply $\rmZ>0$ as explained in Ref.~\cite{Cacciapaglia:2006pk}.} The scale $\Lambda_{\textrm{max}}(\rmZ)\equiv m_W/\sqrt{Z}$ represent the {\it{maximal}} cutoff associated to Eq.~\ref{LZ}: depending on the specific UV completion, the NP is expected to appear at or before $\Lambda_{\textrm{max}}$.\footnote{It is interesting to note that when the equation of motion for the gluon field is used to rewrite Eq.~\ref{LZ} as a four-fermion operator (see Eq.~\ref{Z4f}), the naive cutoff of the EFT, defined as the energy scale at which the $qq\to qq$ amplitude becomes $O(16\pi^2)$, is larger than $\Lambda_{\textrm{max}}$ by a factor $4\pi/g_s$. There is no contradiction, as Eq.~\ref{Z4f} only captures the physics of Eq.~\ref{LZ} at leading order.} Note that $\Lambda_{\textrm{max}}(\rmZ)$ is only an approximate definition of the maximal scale at which the EFT will break down.  
A more precise description requires specifying both the UV theory replacing Eq.~\ref{LZ}, and the value of the required accuracy.

Since we are using high energy jet data to extract bounds on Eq.~\ref{LZ}, we have to make sure that the EFT we are using is under control. Studies along this direction are already present in the literature in the framework of the SM EFT and DM production at colliders~\cite{Pobbe:2017wrj,Goodman:2011jq,Shoemaker:2011vi,Busoni:2013lha,Profumo:2013hqa,Busoni:2014sya,Busoni:2014haa,Matsumoto:2014rxa,Racco:2015dxa,Matsumoto:2016hbs,Contino:2016jqw,Farina:2016rws}. We adopt the following simple  procedure. For a given analysis we recalculate the bound on $\rmZ$ including only events for which the relevant energy variable, $m_{jj}$ or $p_T$, is below a given value $\Lambda_{\textrm{cut}}$. The 95\% CL bound now becomes  a function of $\Lambda_{\textrm{cut}}$, $\rmZ(\Lambda_{\textrm{cut}})$. This bound will be consistent with the validity of the EFT if 
%\be
%\Lambda_{\textrm{cut}}>\Lambda_{\textrm{max}}(\rmZ(\Lambda_{\textrm{cut}}))
%\ee
%or equivalently
\be\label{eftvalidityeq}
\rmZ(\Lambda_{\textrm{cut}})\lesssim\frac{m_W^2}{\Lambda^2_{\textrm{cut}}}\,,
\ee
when using events with $m_{jj}<\Lambda_{\textrm{cut}}$ and
\be\label{eftvalidityeq1}
\rmZ(\Lambda_{\textrm{cut}})\lesssim\frac{m_W^2}{4\Lambda^2_{\textrm{cut}}}\,,
\ee
when using events with $p_T<\Lambda_{\textrm{cut}}$. The factor 1/4 appearing in Eq.~\ref{eftvalidityeq1} is chosen because for a given invariant mass $m_{jj}$ the maximal available $p_T$ is $m_{jj}/2$.

%%%%%%%%%%%%%
%%%%%%%%%%%%%
%%%%%%%%%%%%%
%%%%%%%%%%%%%
\begin{figure}[t]
\begin{center}
\includegraphics[width=0.472\textwidth]{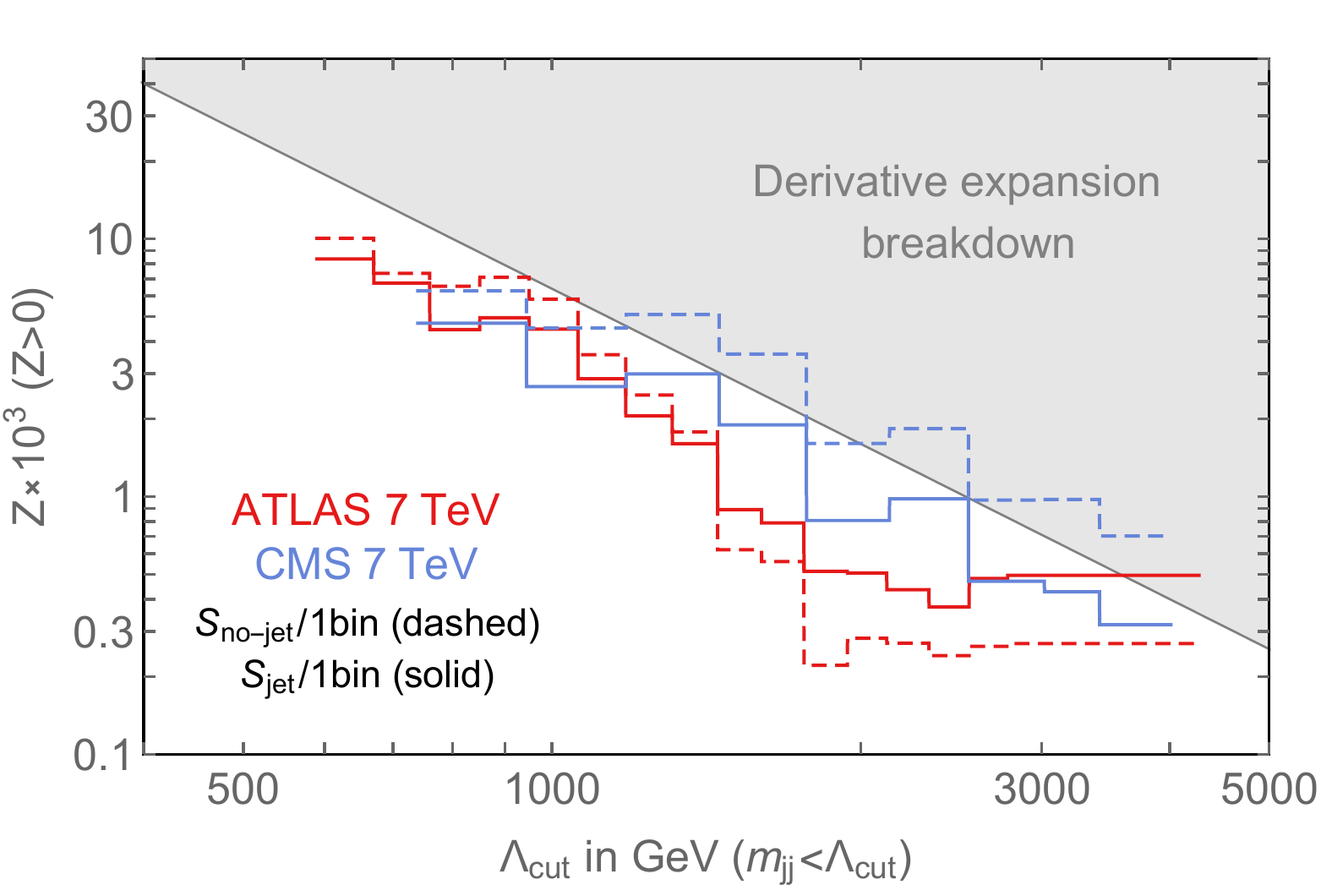}~~~~~~\includegraphics[width=0.464\textwidth]{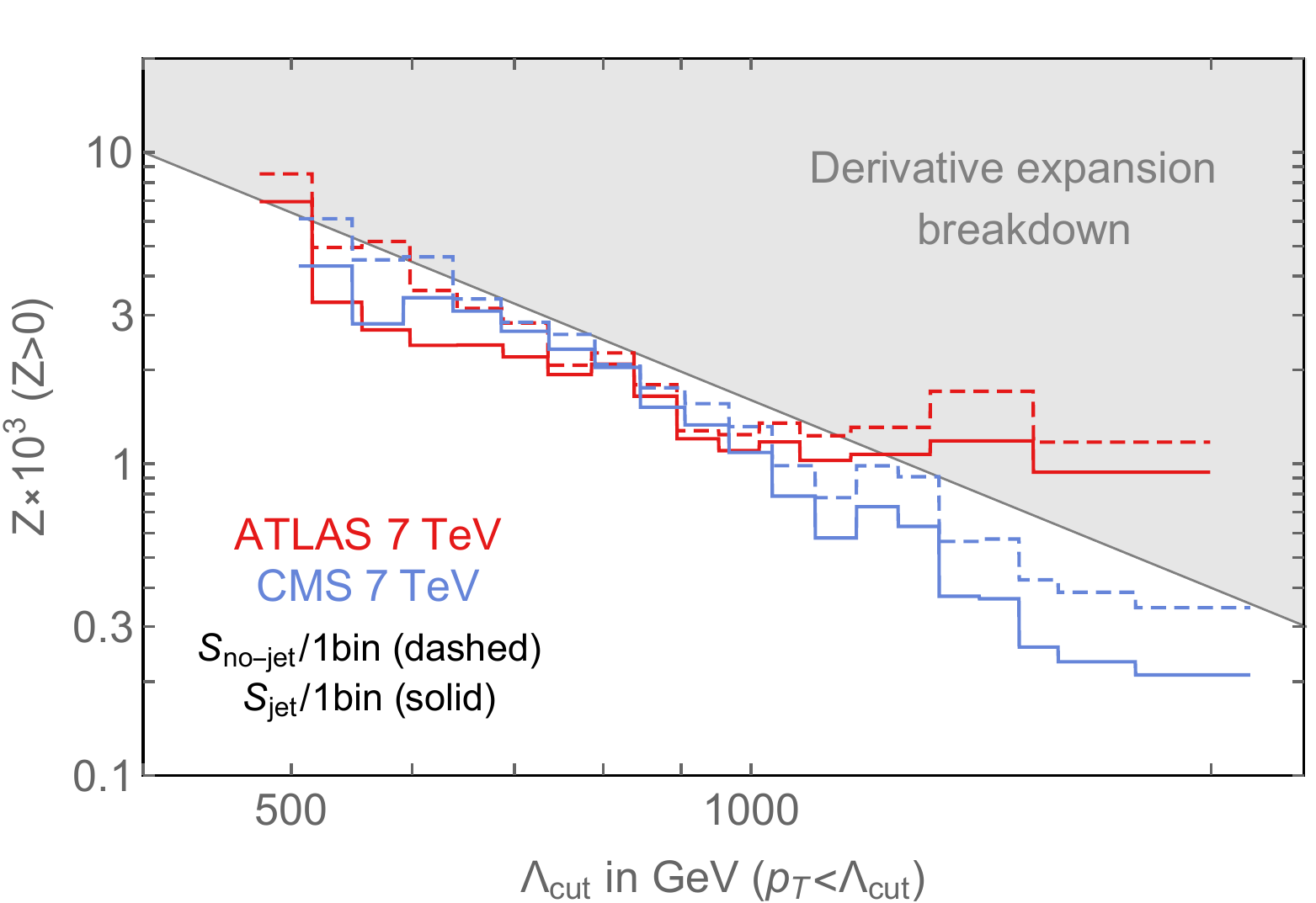}
\end{center}
\vspace{-.3cm}
\caption{ \footnotesize
95\% CL bounds on $\rmZ$ from the 7\,TeV ATLAS and CMS analyses described in the text, as a function of $\lambdacut$. For a given value of $\lambdacut$ only events with $m_{jj}<\lambdacut$ for dijets (left panel) or $p_T<\lambdacut$ for inclusive jets (right panel), are used to extract the limit. The gray area is defined as $\rmZ> m_W^2/\lambdacut^2$ for dijets and $\rmZ> m_W^2/(4\lambdacut^2)$ for inclusive jets, and roughly corresponds to the region in which the EFT description is no longer a valid approximation (see Eq.~\ref{eftvalidityeq}).
}
\label{lambdacut7}
\end{figure}
%%%%%%%%%%%%%
%%%%%%%%%%%%%
%%%%%%%%%%%%%
%%%%%%%%%%%%%

In Fig.~\ref{lambdacut7}, we show the 95\% CL bounds on $\rmZ$, for $\rmZ>0$, as a function of $\Lambda_{\textrm{cut}}$ for the 7\,TeV LHC analyses. We see, as expected, that the bound degrades lowering the value of $\lambdacut$. We also see how the large uncertainties associated to the PDF set $\mathcal S_{\textrm{no-jet}}$ make the limits barely consistent with the validity of the EFT\@. On the other hand the bounds extracted using $\mathcal S_{\textrm{jet}}$ lie, for most cases, below the shaded region in the Figure, representing the bound in Eq.~\ref{eftvalidityeq}.%\footnote{When displaying the bounds extracted from the inclusive jet analyses, the region corresponding to the breakdown of the EFT validity is defined by $\rmZ>m_W^2/(4\lambdacut^2)$} 
A quantitatively similar result holds if, in place of $\mathcal S_{\textrm{jet}}$, we use $\mathcal S_{\textrm{no-jet}}$ and fit the two most central rapidity bins.

\begin{figure}[t]
\begin{center}
\includegraphics[width=0.5\textwidth]{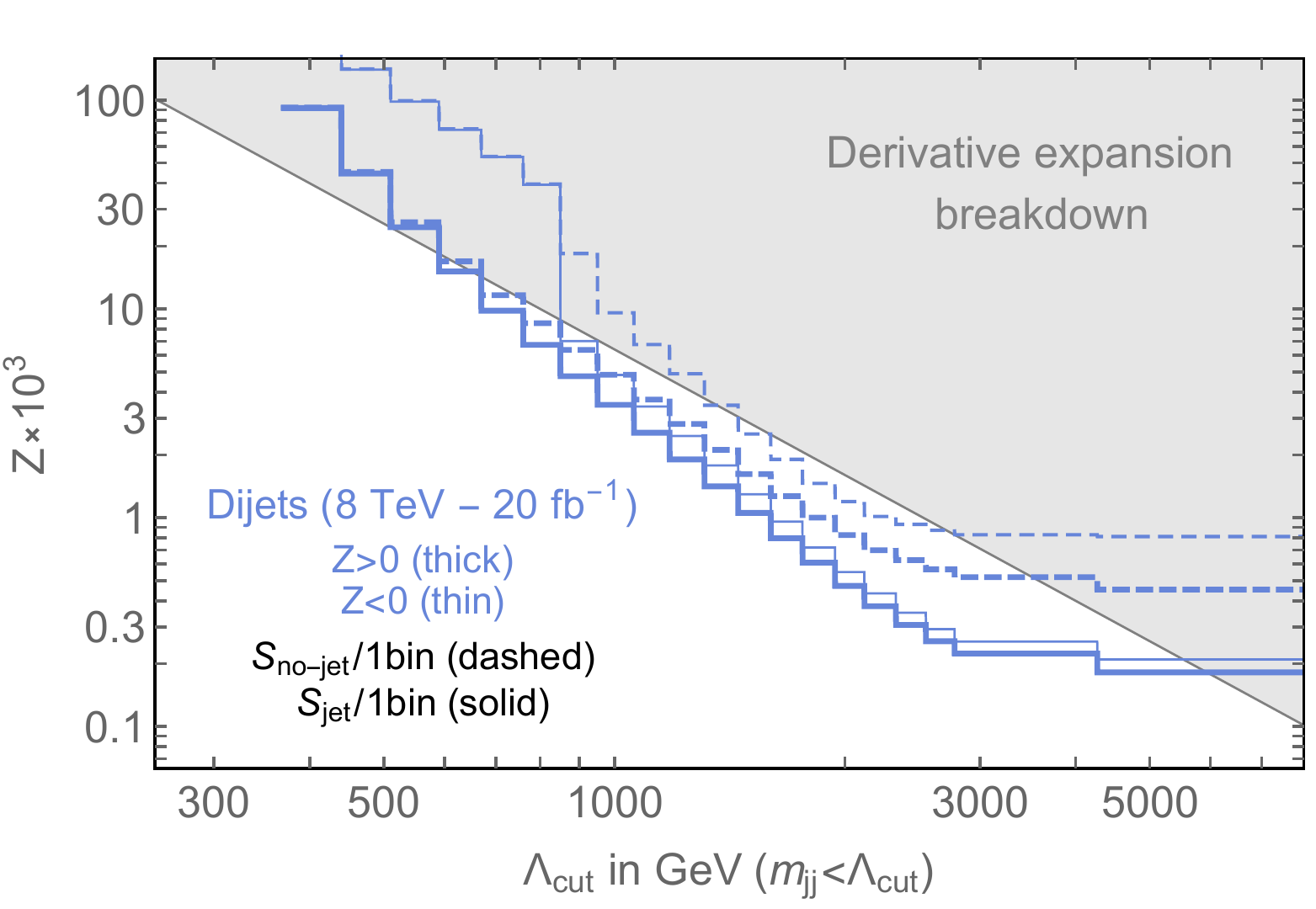}
\end{center}
\vspace{-.3cm}
\caption{ \footnotesize
$\lambdacut$ plot for the 8\,TeV dijet projection. We use only the central rapidity bin, $y^*<0.5$, to extract the limit.  We show the different behavior between PDF sets $\mathcal S_{\textrm{no-jet}}$ and $\mathcal S_{\textrm{jet}}$. Fitting the first two rapidity bins of the dijet distribution yields quantitatively similar results to the $\mathcal S_{\textrm{jet}}$ curve. 95\% CL limits are shown for both $\rmZ>0$ (constructive interference with the SM) and $\rmZ<0$ (destructive interference with the SM).
}
\label{lambdacut8}
\end{figure}

The $\lambdacut$ plot for the 8\,TeV dijet projection is shown in Fig.~\ref{lambdacut8}. Again, while the bound extracted from $\mathcal S_{\textrm{no-jet}}$ is barely consistent with the validity of the EFT, the one obtained from $\mathcal S_{\textrm{jet}}$ is well within the allowed region.

\begin{figure}[t]
\begin{center}
\includegraphics[width=0.465\textwidth]{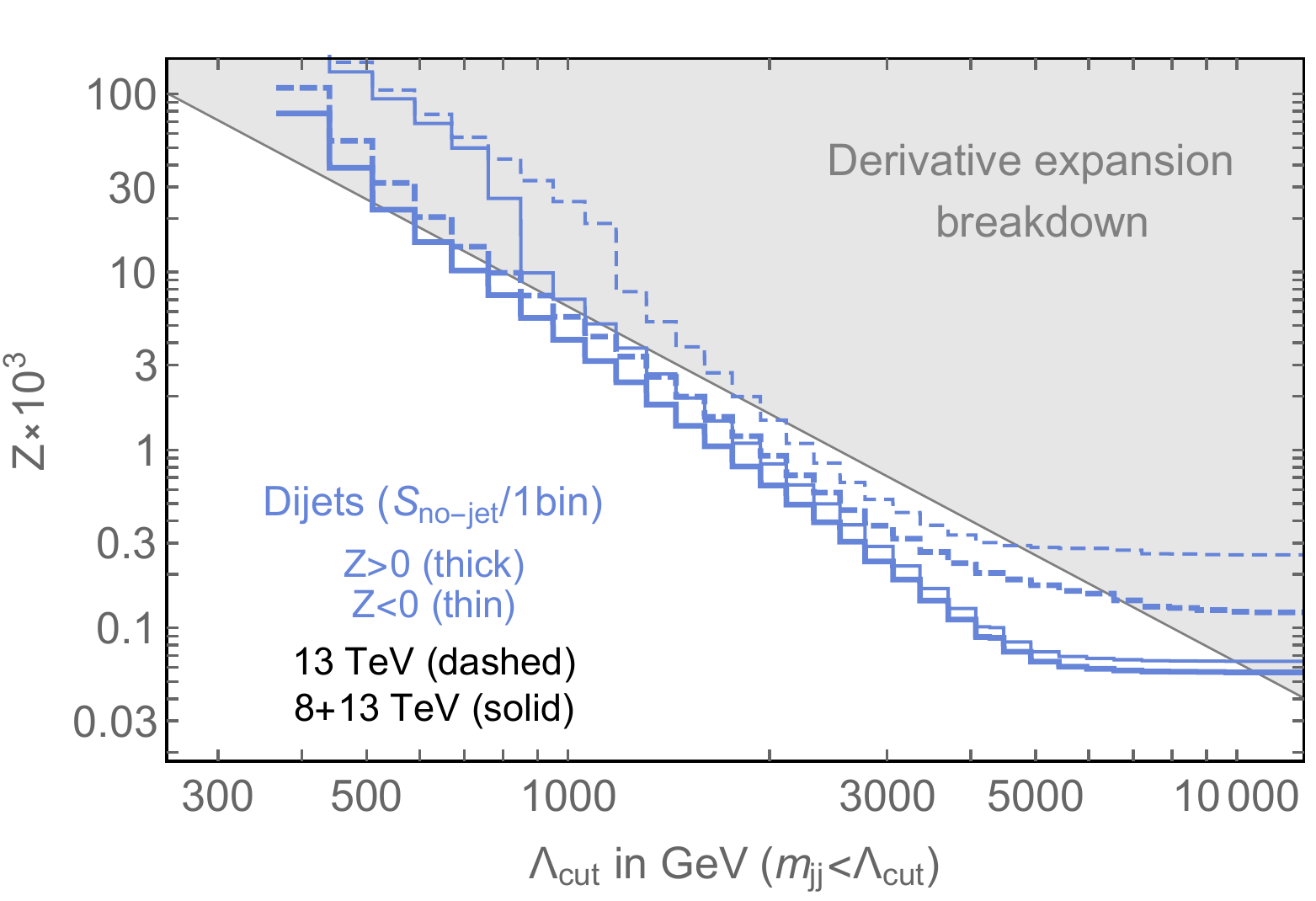}~~~~~~\includegraphics[width=0.485\textwidth]{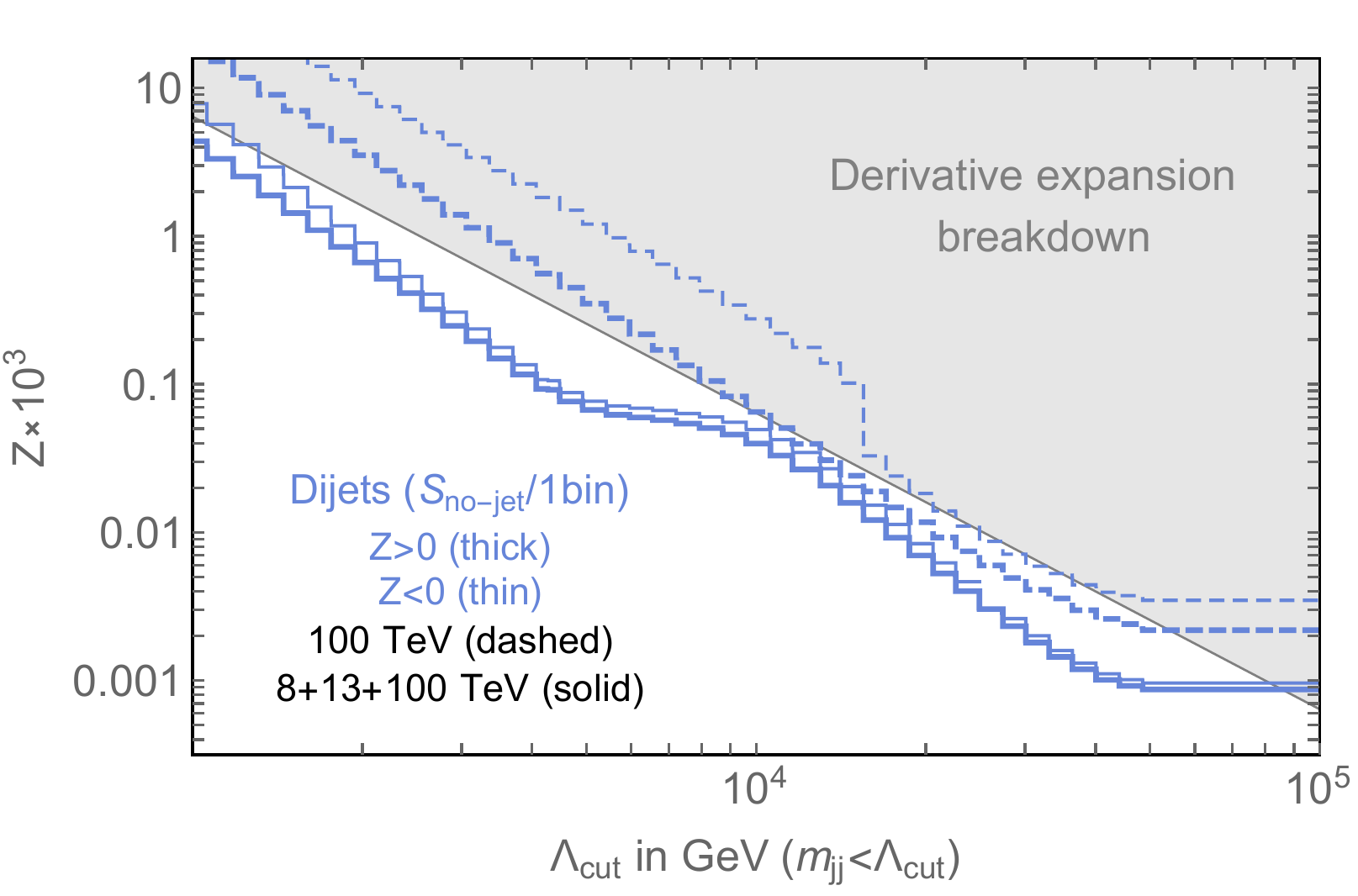}
\end{center}
\vspace{-.3cm}
\caption{ \footnotesize
$\lambdacut$ plot for the projected 95\% CL limits at 13\,TeV and a future 100\,TeV circular $pp$ collider, extracted from dijet searches. The dashed lines show the results of the fit using $\mathcal S_{\textrm{no-jet}}$ and the central rapidity bin, only at a given center of mass energy. We assume an integrated luminosities of $300\,{\textrm{fb}}^{-1}$ at 13\,TeV, and $10\,{\textrm{ab}}^{-1}$ at 100\,TeV\@. The solid lines show the effect of combining different center of mass energies to constrain the PDFs. The combination assumes $20\,{\textrm{fb}}^{-1}$ at 8\,TeV and $3\,{\textrm{ab}}^{-1}$ for the 13\,TeV dataset in the 8+13+100\,TeV combination. The bounds are almost insensitive to the choice of the 13\,TeV integrated luminosity, as explained in the text.
}
\label{lambdacutcombo}
\end{figure}

Finally, in Fig.~\ref{lambdacutcombo} we show the $\lambdacut$ dependence of our 13\,TeV and 100\,TeV projections. In this case, we use $\mathcal S_{\textrm{no-jet}}$, but combine 8+13\,TeV and 8+13+100\,TeV results in order to constrain the PDF variations. For the 100\,TeV case, Fig.~\ref{lambdacutcombo} shows the effect of the 13+100\,TeV combination: lowering $\lambdacut$ below 10\,TeV pushes the bound on $\rmZ$, extracted from the 100\,TeV data alone, outside the validity of the EFT.  The inclusion of the 13\,TeV dataset brings the bound back into the allowed region.
Similar results, which are not shown here, can be obtained from our inclusive jet projections.

 Note that the energy-combination bounds fall more steeply than the naive expectation of $\lambdacut^{-2}$. The reason for this is the different energy dependence of the gluon parton luminosity (which controls the background) and the valence quark one (which controls the signal), the latter falling more slowly with energy than the former.

These $\lambdacut$ plots have many interesting features. They show, for instance, which energy scales are responsible for setting the bound on $\rmZ$. As $\lambdacut$ is increased, each bound improves up to a certain $\lambdacut^*$, above which the bound flattens out. This is the energy at which the suppression due to the PDF luminosity becomes too small to overcome the energy growth due to $\rmZ$. We see that for dijets, this energy scale corresponds to roughly $\lambdacut^* \sim 3$\,TeV at $\sqrt s = 8$\,TeV, and $\lambdacut^* \sim 5, 50$\,TeV for center of mass energies of $13$ and 100\,TeV, respectively.

As we will demonstrate in the next Section, our $\lambdacut$-dependent bound on $\rmZ$ can be used to constrain explicit NP models generating Eq.~\ref{LZ} at low energy. If $M$ is the mass scale at which new states are present, one can identify $\lambdacut = c\times M$, with $c\lesssim 1$ and use $\rmZ<\rmZ(\lambdacut)$ to constrain $M$.

%%%%%%%%%%%%%
%%%%%%%%%%%%%
%%%%%%%%%%%%%
%%%%%%%%%%%%%
%%%%%%%%%%%%%
%%%%%%%%%%%%%
%%%%%%%%%%%%%
%%%%%%%%%%%%%

%%%%%%%%%%%%%
%%%%%%%%%%%%%
%%%%%%%%%%%%%
%%%%%%%%%%%%%
%%%%%%%%%%%%%
%%%%%%%%%%%%%
%%%%%%%%%%%%%
%%%%%%%%%%%%%

%%%%%%%%%%%%%%%%%%%%%%%%%%%%%%%%%%%%%%%%%%%%%%%%%%%%%%%%%%%%%%%%%%%%%%%%%%
\section{Constraints on a vector color octet}
\label{sec:vec}
%%%%%%%%%%%%%%%%%%%%%%%%%%%%%%%%%%%%%%%%%%%%%%%%%%%%%%%%%%%%%%%%%%%%%%%%%%
A simple and motivated extension of the SM, that can be constrained by our bounds, is the theory of a new massive color octet vector boson, $\mathcal G$, which couple to the SM quark $SU(3)_c$ current~\cite{Hall:1985wz,Frampton:1987dn,Frampton:1987ut,Bai:2010dj,Dobrescu:2013coa}
\be\label{Loctet}
\Delta \mathscr L_{\mathcal G}=-\frac{1}{4}D_{[\mu}\mathcal G_{\nu ]}^A D^{[\mu}\mathcal G^{\nu] A}+\frac{M^2}{2}\mathcal G^A_\mu \mathcal G^{A\mu}-g_{\mathcal G} \mathcal G^A_\mu\sum_q \bar q\gamma^\mu T^A q.
\ee
where $D_\mu$ is the $SU(3)_c$ covariant derivative for the adjoint representation, and $T^A$ are the $SU(3)_c$ generators in the fundamental representation.
The model is described by two parameters: the mass $M$ of the new color octet and its coupling $g_{\mathcal G}$ to the quark current. The Lagrangian in Eq.~\ref{Loctet} has to be considered an effective description below the energy scale $\Lambda=(4\pi/g_{\mathcal G}) M$, at which $\mathcal G$ self-interactions become strong. Eq.~\ref{Loctet} can emerge as the low energy description of a composite theory in which $\mathcal G$ represents an excited state of the gluon or, for instance, as the Lagrangian describing the interactions of the lightest Kaluza-Klein partner of the gluon in an extra-dimensional theory. 

It is worth pointing out the existence of simple weakly coupled UV completions of Eq.~\ref{Loctet}. Consider an extension of the SM in which the $SU(3)_c$ color group is replaced by an $SU(3)_1\times SU(3)_2$ gauge theory with gauge couplings $g_1$ and $g_2$. We take the SM quarks to transform under $SU(3)_1$. Introducing a scalar $\Phi$, transforming as a bi-fundamental under the gauge group, and assuming $\Phi$ obtains a vacuum expectation value $\langle\Phi\rangle=V\times I_{3\times 3}$, the gauge group is broken down to $SU(3)_{1+2}$, which is identified with SM color. The orthogonal combination obtains a mass from the kinetic term of $\Phi$,
\be
{\textrm Tr} (D_\mu \Phi^\dagger D^\mu\Phi)\supset\frac{V^2}{2}(g_1G_1^{\mu A}-g_2G_2^{\mu A})^2.
\ee
The physical states, the gluon $G$ and the heavy octet $\mathcal G$, are given by the linear combinations
\be
G= \frac{g_2G_1^{\mu}+g_1G_2^{\mu}}{\sqrt{g_1^2+g_2^2}}\,,\qquad\mathcal G= \frac{-g_1G_1^{\mu}+g_2G_2^{\mu}}{\sqrt{g_1^2+g_2^2}}\,.
\ee
Matching to Eq.~\ref{Loctet} it follows that
\be
g_s=\frac{g_1g_2}{\sqrt{g_1^2+g_2^2}}\,,\qquad M=V\sqrt{g_1^2+g_2^2}\,,\qquad g_{\mathcal G}=\frac{g_1^2}{\sqrt{g_1^2+g_2^2}}\,.
\ee

When $\mathcal G$ is light enough, it can be singly produced at the LHC through $q \bar q\to \mathcal G$, and observed through its decay to dijets. The experimental signature is a bump in the dijet invariant mass distribution.  On the other hand, as the mass of $\mathcal G$ is increased, these searches are expected to become ineffective.   In this heavy octet limit, integrating out $\mathcal G$ in Eq.~\ref{Loctet} generates, at leading order in $1/M^2$, the effective four-fermion operator
\be
\mathscr L\supset -\frac{g^2_{\mathcal G}}{2M^2}J_\mu^A J^{\mu A}.
\ee
According to Eq.~\ref{Z4f}, this four-fermion operator corresponds to a value of $\rmZ$ given by
\be\label{Zoctet}
\rmZ=\frac{g^2_{\mathcal G}}{g_s^2}\frac{m_W^2}{M^2}.
\ee

%%%%%%%%%%%%%
%%%%%%%%%%%%%
%%%%%%%%%%%%%
%%%%%%%%%%%%%
\begin{figure}[t]
\begin{center}
\includegraphics[width=0.95\textwidth]{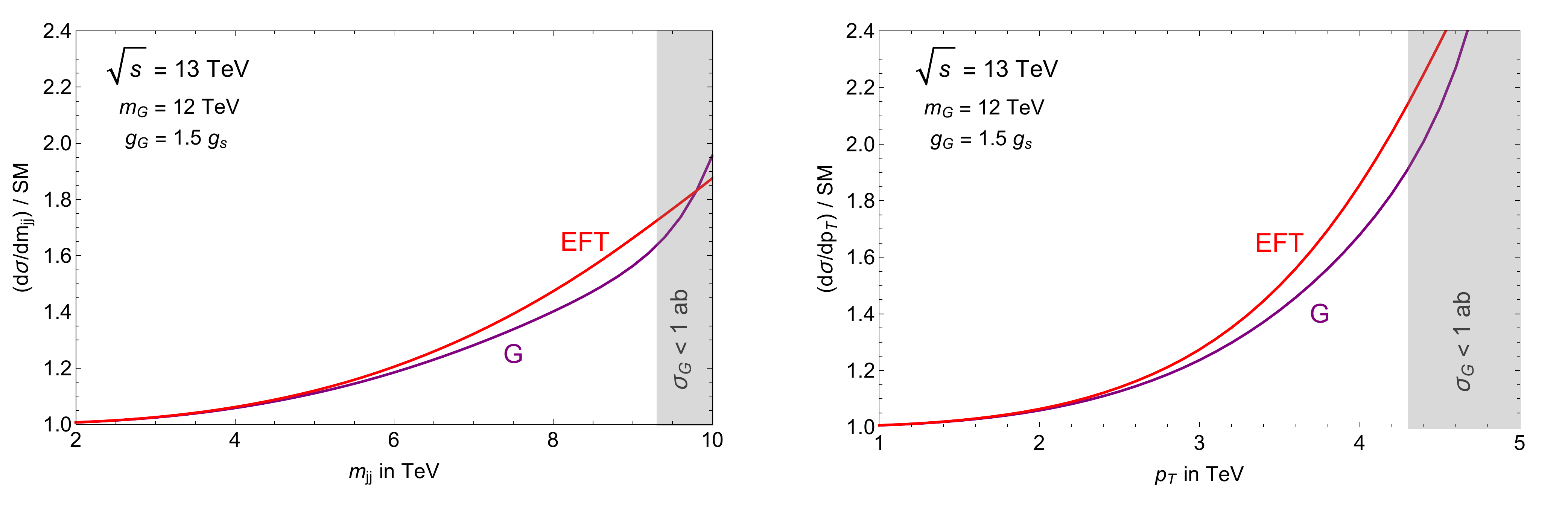}\\\vspace{0.5cm}
\includegraphics[width=0.95\textwidth]{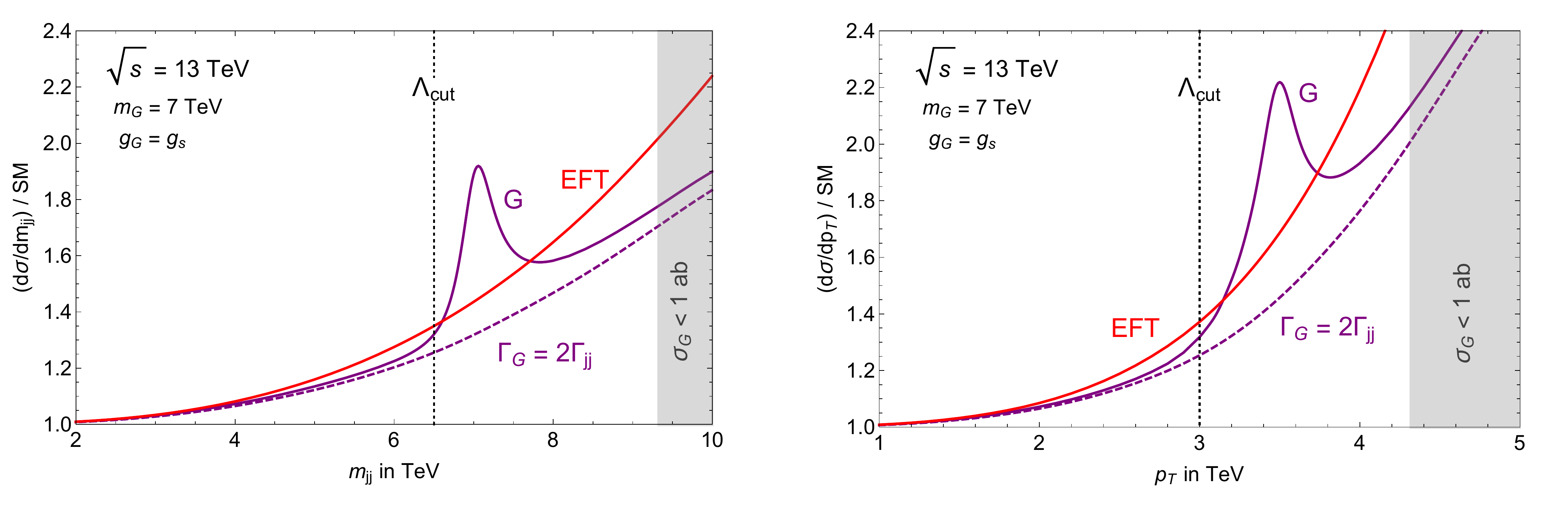}
\end{center}
\vspace{-.3cm}
\caption{ \footnotesize
Dijet (left panels) and inclusive jet (right panels) differential cross sections normalized to SM in the vector octet model for $\sqrt s=13$\,TeV\@. Purple lines show the predictions for $pp\to\mathcal G\to jj$ from Eq.~\ref{Loctet}, while red ones are calculated using the EFT, Eq.~\ref{Zoctet}. In the top row we show a heavy benchmark point, $M=12$\,TeV, in which $\mathcal G$ cannot be produced on-shell. The EFT is in very good agreement with the exact result all the way up to the point at which the cross section become negligibly small, as shown by the gray band. In the bottom row $\mathcal G$ is lighter, $M=7$\,TeV, and we show two possible widths $\Gamma=\Gamma_{\mathcal G}$ and $\Gamma=2\Gamma_{\mathcal G}$. In this case the EFT agrees with the UV completion, as long as the relevant energy variable is below $\lambdacut$, where $\lambdacut=M-\Gamma_{\mathcal G}/2$ for the invariant mass distributions and $\lambdacut=1/2(M-\Gamma_{\mathcal G}/2)$ for the transverse momentum distributions.}
\label{octeteft}
\end{figure}
%%%%%%%%%%%%%
%%%%%%%%%%%%%
%%%%%%%%%%%%%
%%%%%%%%%%%%%

In Fig.~\ref{octeteft}, we show how  $\mathcal G$ affects the dijet and inclusive jet cross sections at the LHC for $\sqrt s=13$\,TeV\@. We compare the calculation of $pp\to\mathcal G\to jj$, using Eq.~\ref{Loctet}, with the EFT prediction from Eq.~\ref{Zoctet}. We use two benchmark points for the mass, the first one, $m_{\mathcal G}=12$\,TeV, in which the octet is too heavy to be directly produced and the second one, $m_{\mathcal G}=7$\,TeV, in which the new vector can be produced on-shell. For this second case we also increase the width of $\mathcal G$ with respect to the value predicted by Eq.~\ref{Loctet}
\be\label{octetwidth}
\Gamma_{\mathcal G}=\frac{g_{\mathcal G}^2}{4\pi} M.
\ee
Fig.~\ref{octeteft} shows that the EFT calculation is able to describe the cross section with better than 10\% accuracy throughout its naive regime of applicability: for the heavy benchmark point this is the whole kinematical range available at 13\,TeV, while for the light benchmark case we take it to be $m_{jj}< M-\Gamma_{\mathcal G}/2$ for the dijet analysis and $2p_{T}< M-\Gamma_{\mathcal G}/2$ for the inclusive jet one.

Conventional bump-hunt searches (see for instance ATLAS~\cite{Aad:2010ae,Aad:2011aj,Aad:2014aqa,ATLAS:2015nsi,ATLAS:2016xiv,ATLAS:2016lvi,Aaboud:2017yvp} and CMS~\cite{Khachatryan:2010jd,Chatrchyan:2011ns,CMS:2012yf,Chatrchyan:2013qha,
Khachatryan:2015sja,Khachatryan:2015dcf,Khachatryan:2016ecr,Sirunyan:2016iap,Sirunyan:2017dnz}) are insensitive to the effects of a dijet resonance that is too heavy to be produced directly. Conversely, a fit to the differential distribution in terms of Eq.~\ref{LZ} or, more generally, the SM EFT, is able to capture the virtual effects of $\mathcal G$ in a theoretically solid, model independent way. Even for a lighter resonance, the tail of low energy events with either $m_{jj}<\lambdacut$ or $p_{T}<\lambdacut$, displayed in the bottom row panels of Fig.~\ref{octeteft}, would be missed by a search looking for a narrow resonance.

%%%%%%%%%%%%%
%%%%%%%%%%%%%
%%%%%%%%%%%%%
%%%%%%%%%%%%%
\begin{figure}[t]
\begin{center}
\includegraphics[width=0.5\textwidth]{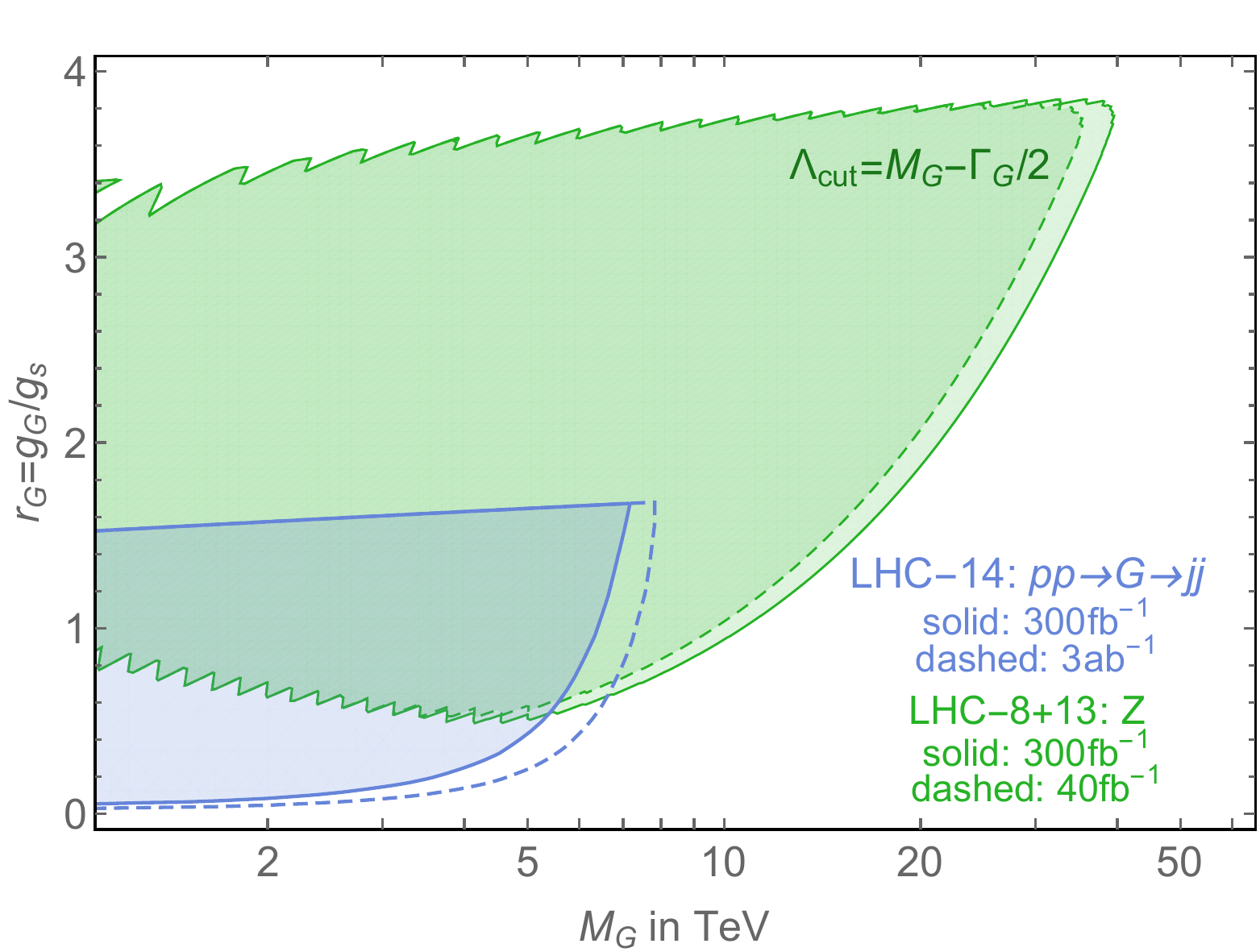}
\end{center}
\vspace{-.3cm}
\caption{ \footnotesize
Comparison of future reach of conventional bump-hunt searches and the EFT limit, for the color octet vector $\mathcal G$. The blue shaded region, from~\cite{Yu:2013wta}, shows the projected limits of 14\,TeV LHC on $pp\to\mathcal G\to jj$ while the green shaded region corresponds to the EFT reach obtained by combining 8\,TeV and 13\,TeV datasets. The limits are obtained by using the 95\% CL bound in Fig.~\ref{lambdacutcombo} with $\lambdacut=M-\Gamma_{\mathcal G}/2$.
}
\label{octetbounds}
\end{figure}
%%%%%%%%%%%%%
%%%%%%%%%%%%%
%%%%%%%%%%%%%
%%%%%%%%%%%%%

We can now compare the projected reach of traditional dijet resonance searches on the parameter space of the model in Eq.~\ref{Loctet}, with our limits on $\rmZ$. We do this in Fig.~\ref{octetbounds}. The LHC limits on $\mathcal G$ are taken from Ref.~\cite{Yu:2013wta}. For the bound on $\rmZ$, we use  the one shown in Fig.~\ref{lambdacutcombo} for the 8+13\,TeV combination. We take $\lambdacut=M-\Gamma_{\mathcal G}/2$. We see that, while at low masses and couplings the resonant searches provide the strongest constraint on the model, their reach degrades both at large $M$ and at large $g_{\mathcal G}$, as $\mathcal G$ becomes a wide resonance. Conversely, the EFT bounds are able to access a much wider region in masses and coupling, up to 30\,TeV for $g_{\mathcal G}\approx 3$. Notice that even for a vector resonance which is as weakly coupled as QCD, $g_{\mathcal G}=g_s$, the EFT bound extends up to 10\,TeV\@. As already noticed, even a large increase in luminosity, going from $40\,{\textrm{fb}}^{-1}$ to $300\,{\textrm{fb}}^{-1}$ only results in a modest increase in the reach. This example opens up the exciting possibility of testing, and maybe even discovering, NP within the dataset already available from Run-II of the LHC\@.

%%%%%%%%%%%%%%%%%%%%%%%%%%%%%%%%%%%%%%%%%%%%%%%%%%%%%%%%%%%%%%%%%%%%%%%%%%
\section{Other operators}
\label{sec:otherops}
%%%%%%%%%%%%%%%%%%%%%%%%%%%%%%%%%%%%%%%%%%%%%%%%%%%%%%%%%%%%%%%%%%%%%%%%%%
In this Section, we comment on other dimension 6 operators that can affect jet physics at high energies. Requiring that there is at least one $2\to 2$ SM amplitude with which the dimension 6 operator can interfere at tree-level, generating an effect that grows with energy, all the relevant terms can be written as a linear combination of four quarks operators.
Assuming that the NP satisfies Minimal Flavor Violation~\cite{DAmbrosio:2002vsn}, the full set of four quark operators has been classified by Ref.~\cite{Domenech:2012ai}. Neglecting terms suppressed by the Yukawa couplings, or involving nontrivial flavor contractions, the list of operators is the following:

\begin{subequations}\label{ops}
  \begin{alignat}{2}
&\mathcal O_{qq}^{(1)}=\frac{1}{2}(\bar q_L\gamma^\mu q_L)(\bar q_L\gamma_\mu q_L)&&\qquad \mathcal O_{qq}^{(8)}=\frac{1}{2}(\bar q_L\gamma^\mu T^Aq_L)(\bar q_L\gamma_\mu T^Aq_L)\\
&\mathcal O_{qu}^{(1)}=(\bar q_L\gamma^\mu q_L)(\bar u_R\gamma_\mu u_R)&&\qquad \mathcal O_{qu}^{(8)}=(\bar q_L\gamma^\mu T^Aq_L)(\bar u_R\gamma_\mu T^Au_R)\\
&\mathcal O_{qd}^{(1)}=(\bar q_L\gamma^\mu q_L)(\bar d_R\gamma_\mu d_R)&&\qquad \mathcal O_{qd}^{(8)}=(\bar q_L\gamma^\mu T^Aq_L)(\bar d_R\gamma_\mu T^Ad_R)\\
&\mathcal O_{uu}^{(1)}=\frac{1}{2}(\bar u_R\gamma^\mu u_R)(\bar u_R\gamma_\mu u_R)&&\qquad \mathcal O_{uu}^{(8)}=\frac{1}{2}(\bar u_R\gamma^\mu T^Au_R)(\bar u_R\gamma_\mu T^Au_R)\\
&\mathcal O_{ud}^{(1)}=(\bar u_R\gamma^\mu u_R)(\bar d_R\gamma_\mu d_R)&&\qquad \mathcal O_{ud}^{(8)}=(\bar u_R\gamma^\mu T^Au_R)(\bar d_R\gamma_\mu T^Ad_R)\\
&\mathcal O_{dd}^{(1)}=\frac{1}{2}(\bar d_R\gamma^\mu d_R)(\bar d_R\gamma_\mu d_R)&&\qquad\mathcal O_{dd}^{(8)}=\frac{1}{2}(\bar d_R\gamma^\mu T^Ad_R)(\bar d_R\gamma_\mu T^Ad_R) \\
&\mathcal O_{qq}^{(3)}=\frac{1}{2}(\bar q_L\gamma^\mu \tau^a q_L)(\bar q_L\gamma_\mu \tau^a q_L)  \\\nn
  \end{alignat}
  \end{subequations}
  
In Eq.~\ref{ops}, the operators on the right involve a color octet contraction, while those on the left a singlet one.  In order to fix the notation we normalize the operators in Eq.~\ref{ops} according to
\be\label{opsL}
\mathscr L= \sum_I\frac{c_I}{v^2}\mathcal O_I.
\ee
where $v\approx 246$\,GeV is the Higgs vacuum expectation value.

As mentioned in the Introduction, the operator in Eq.~\ref{LZ} is a linear combination of the octet operators in Eq.~\ref{ops}. In particular
\be\label{Z4foctet}
-\frac{\rmZ}{2 m_W^2}(D_\mu G^{\mu\nu A})^2=-\frac{\rmZ g_s^2 }{m_W^2}\sum_{I}\mathcal O_{I}^{(8)}.
\ee

Including all the operators in Eq.~\ref{ops} modifies the dijet/inclusive jet cross section in a given bin in a way analogous to Eq.~\ref{sigmaZ},
\be\label{sigmaallops}
\sigma^{\textrm{th}}_i=\sigma_i^{SM}+\sum_I c_I\sigma_{i\,I}^{(1)} +\sum_{IJ}c_Ic_J \sigma_{i\,IJ}^{(2)},
\ee
where the $c_I$ are the coefficients of the dimension 6 operators in Eq.~\ref{ops}, both singlets and octets. We could now in principle repeat our fits using Eq.~\ref{sigmaallops} to construct a likelihood function, analogously to Eq.~\ref{llh2}. Due to the suppressed size of the SM interference of many combinations of operators in Eq.~\ref{ops}, the resulting likelihood function would be highly non-Gaussian and the result of the global fit would be impossible to present in a closed form. For this reason, instead of attempting to be fully general, in the following we present 8 and 13\,TeV projections for the bounds on a few motivated combinations of operators in Eq.~\ref{ops}.

%%%%%%%%%%%%%
%%%%%%%%%%%%%
%%%%%%%%%%%%%
%%%%%%%%%%%%%
\begin{table}[t]
\caption*{95\%~CL bounds on $\rmW\times 10^4$}
\begin{center}\vspace{-0.5cm}
{\small
\begin{tabular}{c|c|c} 
Analysis & $8$\,TeV ($20\,{\textrm{fb}}^{-1}$)  & $13$\,TeV ($300\,{\textrm{fb}}^{-1}$)\\ \hline \hline
$pp\to jj$  & [-39,+17]  & [-7.0,4.3]   \\ 
$pp\to\ell\ell$  &[-3,+15]$_{{{ATLAS}}}$ / [-5,+22]$_{CMS}$  & [-1.5,+1.5] \\
$pp\to \ell\nu$ & [-3.9,+3.9]&[-0.7,+0.7]
\end{tabular}
} 
\vspace{0.3cm}
\caption{\label{resultsW}\footnotesize 95\% CL projected bounds on the $\rmW$ parameter, extracted from the dijet invariant mass distribution at 8 and 13\,TeV\@. We compare the dijet reach with the neutral and charged Drell-Yan bounds from Ref.~\cite{Farina:2016rws}.  Note that the 8\,TeV $pp\to\ell\ell$ entry corresponds to the recasting of actual data from ATLAS and CMS~\cite{Farina:2016rws}, whereas the other 8 TeV bounds are projections.}
\end{center}
\end{table}
%%%%%%%%%%%%%
%%%%%%%%%%%%%
%%%%%%%%%%%%%
%%%%%%%%%%%%%

%%%%%%%%%%%%%%%%%%%%%%%%%%%%
%%%%%%%%%%%%%%%%%%%%%%%%%%%%
%%%%%%%%%%%%%%%%%%%%%%%%%%%%
\subsection{The {\textrm{W}} parameter}
The $\rmW$ parameter is an oblique correction to the electroweak sector of the SM~\cite{Barbieri:2004qk} induced by the following dimension 6 operator
\be\label{W}
\Delta \mathscr L_{\rmW}=-\frac{\rmW}{2m_W^2}(D_\mu W^{a\,\mu\nu})^2\,.
\ee
Notice the analogy with the definition of $\rmZ$ in Eq.~\ref{LZ}. Using the equation of motion of the $W$ boson, Eq.~\ref{W} corresponds to
\be\label{4fW}
\Delta \mathscr L_{\rmW}=-\frac{g_2^2\rmW}{2m_W^2}(\bar q_L\gamma^\mu \tau^a q_L)(\bar q_L\gamma_\mu \tau^a q_L)=-\frac{g_2^2\rmW}{m_W^2}\mathcal O_{qq}^{(3)}\, .
\ee
The bounds on $\rmW$ we obtain using 8 and 13\,TeV projections for the dijet invariant mass distribution are shown in Table~\ref{resultsW}. Similar bounds are obtained using the  inclusive jet $p_T$ distribution. The bounds on $\rmW$ in Table~\ref{resultsW} are weaker than those on $\rmZ$ extracted from dijets. This is due to the $g_2^2/g_s^2$ suppression of the operator in Eq.~\ref{4fW} compared to Eq.~\ref{Z4f}. 

We find it useful to compare these bounds on $\rmW$ with the bounds obtained in Ref.~\cite{Farina:2016rws}, studying neutral and charged Drell-Yan processes at the LHC.  We find that leptonic final states are superior in constraining $\rmW$, which is a consequence of the smaller systematic uncertainties impacting Drell-Yan cross sections.

%%%%%%%%%%%%%%%%%%%%%%%%%%%%
%%%%%%%%%%%%%%%%%%%%%%%%%%%%
%%%%%%%%%%%%%%%%%%%%%%%%%%%%
\subsection{Quark compositeness}

Composite Higgs models with partial compositeness, which avoid the flavor problem by imposing large flavor symmetries in the strong sector, may require some of the light SM quark chiralities to be strongly coupled to the composite sector in order to explain the top quark mass~\cite{Redi:2013eaa}. If one of the SM quark chiralities is composite one expects operators in Eq.~\ref{ops} to be present in the low energy theory with coefficients
\be
c_I\sim \frac{g_c^2 v^2}{\Lambda_c^2}\, ,
\ee
where $\Lambda_c$ and $g_c$ are the scale of compositeness and the coupling of the SM quarks to the strongly interacting sector, respectively. In Table~\ref{quarkcomp} we show the bounds on four of these operators, assuming that only one of them at a time is present in the Lagrangian. This is a reasonable assumption since, if only one of the chiralities is composite, the operators which are shown are expected to be the most important ones. Notice that we don't include $\mathcal O_{qq}^{(3)}$, $\mathcal O_{uu}^{(8)}$, and $\mathcal O_{dd}^{(8)}$. The reason for this is that, at the LHC, the BSM dijet cross section is dominated by processes which are initiated by  $uu$, $dd$, and $ud$. Keeping only terms involving the first generation, one can show that $\mathcal O_{qq}^{(3)}$ and $\mathcal O_{uu,dd}^{(8)}$ can be rewritten, using Fierz identities, into combinations of operators appearing in Table~\ref{quarkcomp}. The bounds shown in Table~\ref{quarkcomp} are asymmetric as a consequence of suppressed interference with the SM and the importance of the quadratic terms in the cross section (see Eq.~\ref{sigmaallops}). Jet data constrain universal $u_R$ compositeness more strongly than universal $d_R$ compositeness. The reason for this is the larger value of the $u$ quark PDF with respect to the $d$ one. At 13\,TeV, the bounds on the scale $\Lambda_c$ can be as strong as 70\,TeV, for maximal $g_c\sim 4\pi$.

\begin{table}[t]
\caption*{95\%~CL bounds on $c_I \times 10^3$}
\begin{center}\vspace{-0.5cm}
{\small
\begin{tabular}{c|c|c|c}
Scenario & Operator & $8$\,TeV ($20\,{\textrm{fb}}^{-1}$) & $13$\,TeV ($300\,{\textrm{fb}}^{-1}$)\\ \hline \hline 
\multirow{ 2}{*}{composite $q_L$} & $ c_{qq}^{(8)}$ & [-5.4,+10.4] & [-1.3,+2.1] \\ & $ c_{qq}^{(1)}$ & [-2.1,+6.8] & [-0.6,+1.5] \\ \hline
composite $u_R$ & $ c_{uu}^{(1)}$ & [-2.5,+9.1] & [-0.75,+1.8]\\ \hline
composite $d_R$ & $ c_{dd}^{(1)}$ & [-12,+18] & [-3.6,+4.6]\\ 
\end{tabular}
} \vspace{0.3cm}
\caption{\label{quarkcomp}\footnotesize 95\% CL projected bound on the coefficient, $c_I$, of 4 quark operators defined as in Eqs.~\ref{ops} and \ref{opsL}. Our bounds use the 8 and 13\,TeV calculation for the dijet invariant mass distribution. We use only the central rapidity bin and $\mathcal S_{\textrm{jet}}$ in the fit.}
\end{center}
\end{table}

%%%%%%%%%%%%%%%%%%%%%%%%%%%%
%%%%%%%%%%%%%%%%%%%%%%%%%%%%
%%%%%%%%%%%%%%%%%%%%%%%%%%%%
\subsection{The axigluon}
\begin{figure}[t]
\begin{center}
\includegraphics[width=0.45\textwidth]{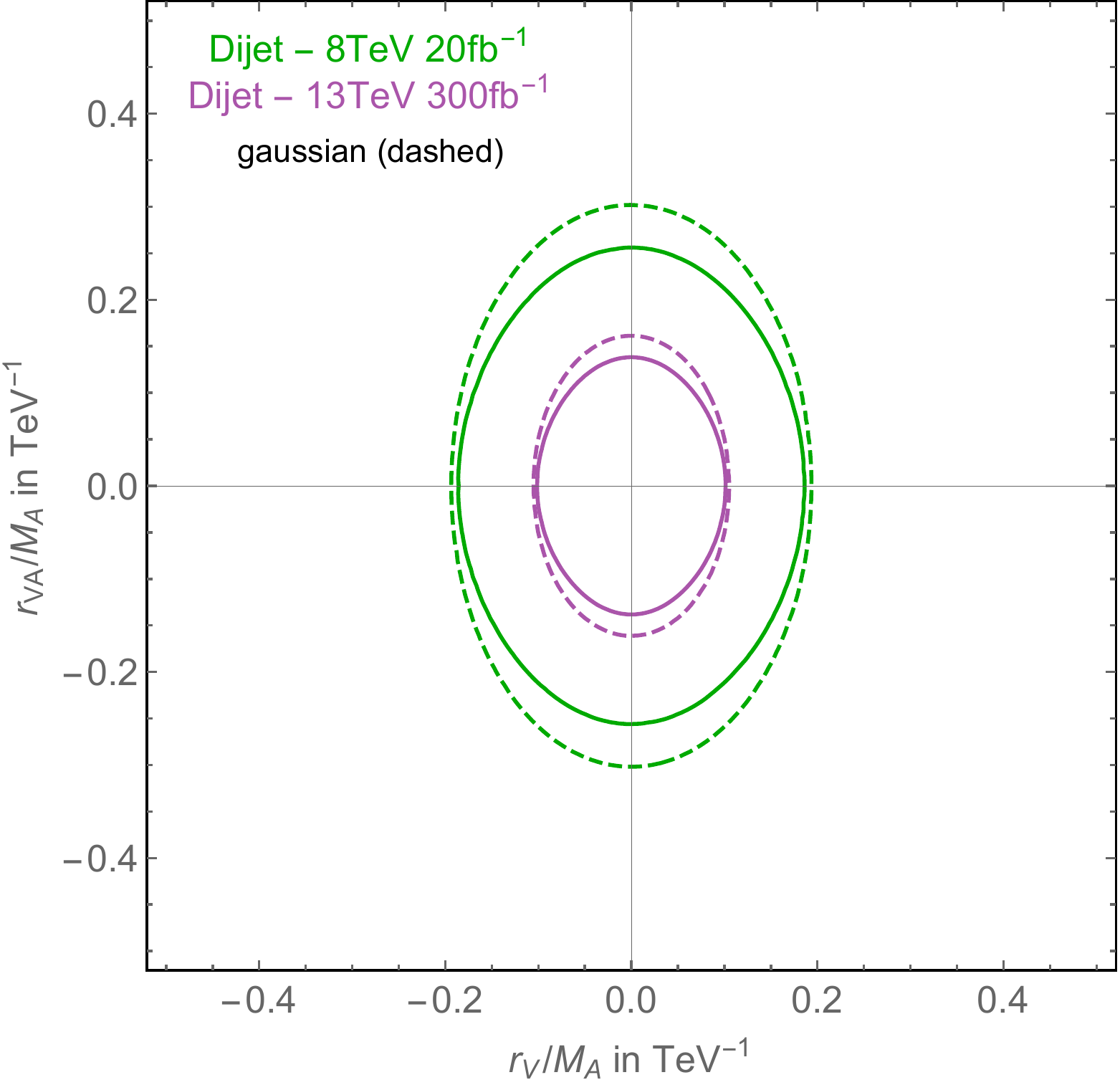}
\end{center}
\vspace{-.3cm}
\caption{ \footnotesize 95\% CL contours for the coefficients $r_V/M_{\mathcal A}$ and  $r_A/M_{\mathcal A}$, in Eq.~\ref{axigluonop}, extracted from our dijets projections at 8 and 13\,TeV\@. We use $\mathcal S_{\textrm{jet}}$, fitting only to the central rapidity bin. The allowed region is the one bounded by the lines. The dashed contours are obtained from the Gaussian approximation to the likelihood function, by discarding quadratic terms from Eq.~\ref{sigmaallops}. 
}
\label{axigluonbounds}
\end{figure}

 A more general scenario than the one presented in Section~\ref{sec:vec} is represented by a color octet vector $\mathcal A$, of mass $M_{\mathcal A}$, whose interactions with quarks is a combination of vector and axial couplings~\cite{Pati:1975ze,Hall:1985wz,Frampton:1987dn,Frampton:1987ut,Bagger:1987fz}.  This so-called axigluon has interactions,
 \be\label{axigluonL}
 \Delta\mathscr L_{\mathcal A}\supset- g_s\mathcal A^A_\mu\sum_q \bar q T^A\gamma^\mu(r_V+r_A\gamma^5) q.
 \ee
Integrating out $\mathcal A$ generates the following dimension 6 operator in the low energy theory:
\be\label{axigluonop}
 \Delta\mathscr L=-\frac{g_s^2}{2 M^2_{\mathcal A}}\left[ \bar q T^A\gamma^\mu(r_V+r_A\gamma^5) q\right]^2.
\ee
In Fig.~\ref{axigluonbounds}, we show the combined bounds bounds on the quantities $r_V/M_{\mathcal A}$ and $r_A/M_{\mathcal A}$, obtained from dijet projection at 8 and 13\,TeV\@. We find that, for $r_V, r_A\sim 1$, the bound on $M_{\mathcal A}$ is of order 5 and 10\,TeV for center of mass energies of 8 and 13\,TeV, respectively.

%%%%%%%%%%%%%
%%%%%%%%%%%%%
%%%%%%%%%%%%%
%%%%%%%%%%%%%
%%%%%%%%%%%%%
%%%%%%%%%%%%%
%%%%%%%%%%%%%
%%%%%%%%%%%%%
%%%%%%%%%%%%%

%%%%%%%%%%%%%%%%%%%%%%%%%%%%%%%%%%%%%%%%%%%%%%%%%%%%%%%%%%%%%%%%%%%%%%%%%%
\section{Conclusions}
\label{sec:conclusions}
%%%%%%%%%%%%%%%%%%%%%%%%%%%%%%%%%%%%%%%%%%%%%%%%%%%%%%%%%%%%%%%%%%%%%%%%%%
In this paper we have used fully differential NLO predictions for the dijet and inclusive jet cross sections, and publicly available data from 7\,TeV LHC,  to constrain the behavior of QCD at high energies.

We focused on the parameter $\rmZ$ defined in Eq.~\ref{LZ} and reproduced here for convenience
\be\nn%\label{LZconclusions}
\Delta \mathscr L= -\frac{\rmZ}{2 m_W^2}(D_\mu G^{\mu\nu A})^2 \, .
\ee
$\rmZ$ can be understood as an oblique correction for the QCD sector of the SM\@. Its effect is to modify the gluon propagator and to induce energy growing terms in $2\to 2$ quark amplitudes. 

%%%%%%%%%%%%%
%%%%%%%%%%%%%
%%%%%%%%%%%%%
%%%%%%%%%%%%%
\begin{figure}[t]
\begin{center}
\includegraphics[width=0.7\textwidth]{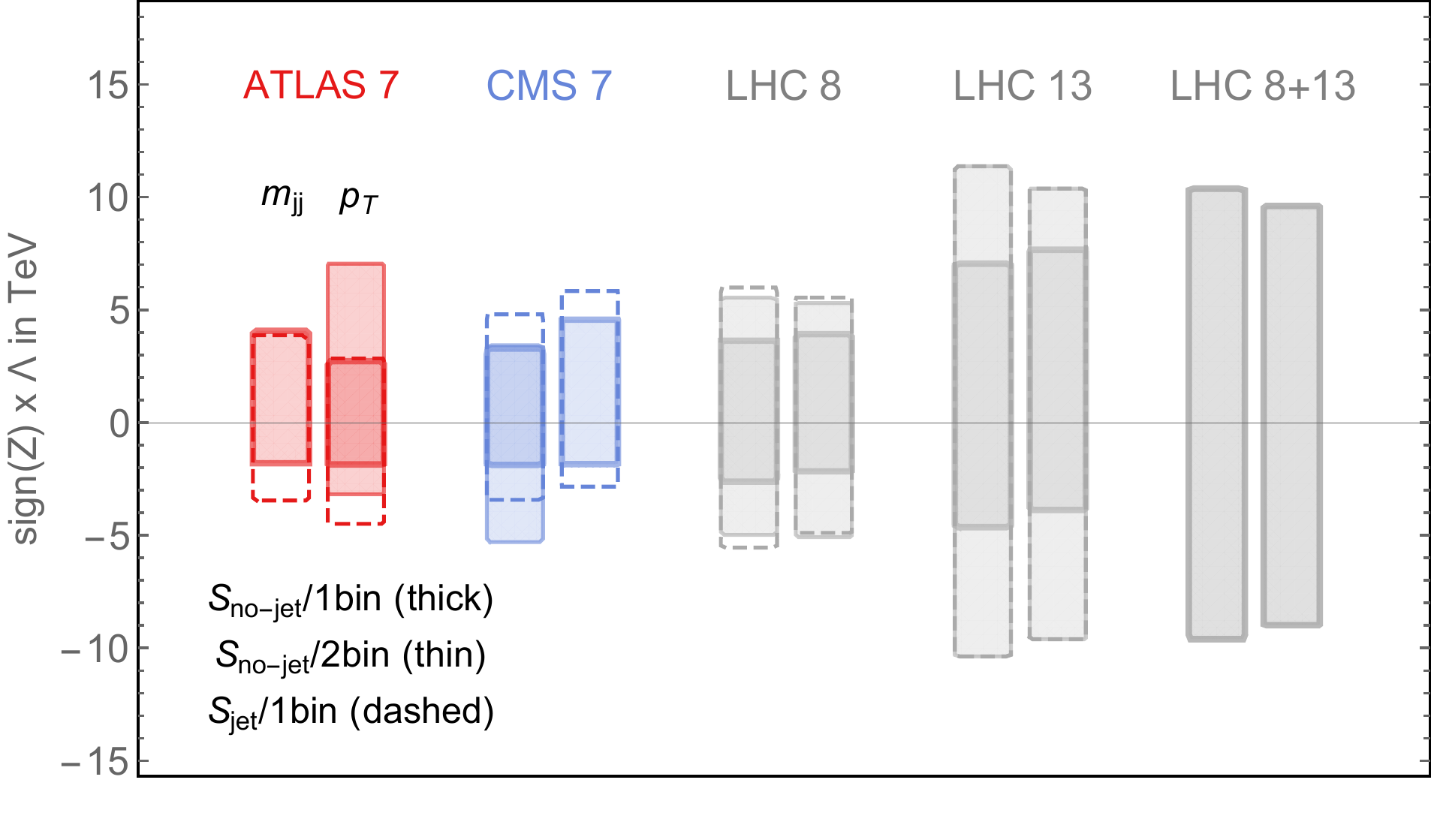}
\end{center}
\vspace{-.3cm}
\caption{ \footnotesize Summary of our results at the LHC, in terms of the NP scale $\Lambda$, defined as $\Lambda\equiv m_W/\sqrt{|\rmZ|}$. For the 8\,TeV LHC we assume an integrated luminosity of $20\,{\textrm{fb}}^{-1}$, while we use $300\,{\textrm{fb}}^{-1}$ for the LHC at 13\,TeV\@.  Red and blue correspond to fits to ATLAS and CMS data, respectively, whereas gray corresponds to future projections at 8 and 13 TeV.}
\label{summaryplot}
\end{figure}
%%%%%%%%%%%%%
%%%%%%%%%%%%%
%%%%%%%%%%%%%
%%%%%%%%%%%%%

The results of our fits to the 7\,TeV data, and projections at 8 and 13\,TeV, are summarized in Fig.~\ref{summaryplot}. We display our projections in terms of the mass scale $\Lambda=m_W/\sqrt{|\rmZ|}$.  We see that dijets and the inclusive jet $p_T$ spectra have similar constraining power. 
The public 7\,TeV data constrain $\Lambda\gtrsim 3.5$\,TeV, and our 8\,TeV projections push this scale to 4$-$5\,TeV\@. 13\,TeV data are expected to double this reach, as shown in the figure. Our projections show, in particular, that the EFT reach quickly saturates: going from $40\,{\textrm{fb}}^{-1}$ to $300\,{\textrm{fb}}^{-1}$ at 13\,TeV only provides a minor increase in the value of $\Lambda$, see Table~\ref{results13}. We also extract the potential reach of a future $pp$ collider, with 100\,TeV center of mass energy, finding a bound on $\Lambda$ of order 80\,TeV\@.

Despite the existence of significant uncertainties of both theoretical and experimental origin, high energy QCD measurements can be a powerful probe of the SM EFT\@. We find, in particular, that our results are competitive with the bounds from Drell-Yan in~\cite{Farina:2016rws} when they can both be applied to the same UV theory. If we assume, for instance, that all the SM gauge bosons are composite at the same energy scale $\Lambda_C$, our 13\,TeV-300\,fb$^{-1}$ dijet projections imply $\Lambda_C\gtrsim 11$\,TeV. This is comparable with the results of~\cite{Farina:2016rws} reporting $\Lambda_C\gtrsim 9.6$\,TeV with the same energy and luminosity, obtained as a constraint on the parameter $\rmW$ from charged Drell-Yan projections.

A crucial element, that enters our fits, is the evaluation of systematic uncertainties of theoretical origin, in particular the scale and the PDF uncertainties. The uncertainty associated to the PDFs is the major limiting factor in constraining $\rmZ$. We discussed three possible ways to deal with this large systematic uncertainty. The first is to use a PDF set including high energy jet data in its global fit (what we call $\mathcal S_{\textrm{jet}}$ in the paper), reducing the uncertainty associated to the gluon PDF\@. The second possibility is to exploit PDF correlations across different rapidity bins by performing a doubly differential fit. Finally, the third option is to exploit PDF correlations across different center of mass energies by performing an 8+13\,TeV data combination. These three different methods give similar results when they are all applicable. In particular, both the doubly differential fit and the energy combination do not suffer from the potential problem of overfitting the data, in case the new physics we are trying to constrain also affects the jet data entering the determination of $\mathcal S_{\textrm{jet}}$.

The possibility to fit Eq.~\ref{LZ} to a differential cross section, either the dijet invariant mass distribution or the inclusive jet transverse momentum distribution, is of fundamental importance. Any EFT is valid only below a certain energy scale, and this scale has to be specified in order to define the theory. The maximal value of the EFT cutoff, for Eq.~\ref{LZ}, is given by $\Lambda_{\textrm{max}}\sim m_W/\sqrt{|Z|}$. However, the actual EFT cutoff can be lower in specific UV completions. Fitting to a differential distribution allows us to assess how the bound on $\rmZ$ changes by lowering this cutoff, and therefore to understand when the EFT framework is useful to describe the data.

We applied the cutoff dependent bounds, shown in Figs.~\ref{lambdacut8} and~\ref{lambdacutcombo}, in the context of a specific UV completion of Eq.~\ref{LZ}, in which $\rmZ$ is generated by integrating out a massive vector boson coupled to the SM $SU(3)_c$ quark current. We find that the EFT is  able to probe a large region of the parameter space of the model, surpassing the mass reach of traditional searches for narrow dijet resonances (see Fig.~\ref{octetbounds}). 

If a deviation from the SM prediction is measured in some high energy observable, then comparing different differential distributions will become of fundamental importance. Given the existence of systematics uncertainties, it is only by comparing multiple distributions and examining their correlations, within the SM EFT framework, that a call for a discovery can be made.

Our results can be extended in various directions. It would be interesting to use the emerging NNLO results of~\cite{Currie:2016bfm,Currie:2017eqf} to reduce the uncertainty associated to the scale variation of the SM predictions. A second improvement in our analysis would be to include NLO BSM effects in our calculation for a more precise estimate of the NP scale. 
Finally, more operators from the SM EFT can be considered. We initiated this program in Section~\ref{sec:otherops}, but have not presented bounds on all possible operators.
We leave these extensions for future work.

%%%%%%%%%%%%%%%%%%%%%%%%%%%%%%%%%%%%%%%%%%%%%%%%%%%%%%%%%%%%%%%%
%%%%%%%%%%%%%%%%%%%%%%%%%%%%%%%%%%%%%%%%%%%%%%%%%%%%%%%%%%%%%%%%%%%%%%%%%%%%%%%%%%%%%%%%%%%%%%%%%%%%%%%%%%%%%%%%%%%%%%%%%%%%%%%%%%%%%%%%%%%%%%%%%%%%%%%%%%%%%%%%%%%

\section*{Acknowledgments}
We thank Kyle Cranmer and Michelangelo Mangano for comments and discussions.  We thank Mario Campanelli, Matthias Schott, and Bogdan Malaescu for useful clarifications about the ATLAS contact operator bounds. M.F. is supported in part by the DOE Grant DE-SC0010008. D.P. and J.T.R. are supported by the NSF CAREER grant PHY-1554858. S.A. acknowledges support by the COFUND Fellowship under grant agreement PCOFUND-GA-2012-600377.
%%%%%%%%%%%%%%%%%%%%%%%%%%%%%%%%%%%%%%%%%%%%%%%%%%%%%%%%%%%%%%%%%%%%%%%%%%%%%%%%%%%%%%%%%%%%%%%%%%%%%%%%%%%%%%%%%%%%%%%%%%%%%%%%%%%%%%%%%%%%%%%%%%%%%%%%%%%%%%%%%%%%%%%%%%%%%%%%%%%%%%%%%%%%%%%%%%%%%%%%%%%%%%%%%%%%%%%%%%%%%%%%%%%%%%%%%%%%%%%%%%%%%%%%%%%%%%%%%%%%%%%%%%%%%%%%%%%%%%%%%%%%%%%%%%%%%%%%

\appendix
%%%%%%%%%%%%%%%%%%%%%%%%%%%%%%%%%%%%%%%%%%%%%%%%%%%%%%%

%%%%%%%%%%%%%%%%%%%%%%%%%%%%%%%%%%%%%%%%%%%%%%%%%%%%%%%%%%%%%%%%%%%%%%%%%%%%%%%%%%%%%%%%%%%%%%%%%%%%%%%%%%%%%%%%%%%%%%%%%%%%%%%%%%%%%%%%%%%%%%%%%%%%%%%%%%%%%%%%%%%%%%%%%%%%%%%%%%%%%%%%%%%%%%%%%%%%%%%%%%%%%%%%%%%%%%%%%%%%%%%%%%%%%%%%%%%%%%%%%%%%%%%%%%%%%%%%%%%%%%%%%%%%%%%%%%%%%%%%%%%%%%%%%%%%%%%%%%%%%%%%%%%%%%%%%%%%%%%%%%%%%%%%%%%%%%%%%%%%%%%%%%%%%%%%%%%%%%

%%%%%%%%%%%%%%%%%%%%%%%%%%%%%%%%%%%%%%%%%%%%%%%%%%%%%%%
%%%%%%%%%%%%%%%%%%%%%%%%%%%%%%%%%%%%%%%%%%%%%%%%%%%%%%%
%%%%%%%%%%%%%%%%%%%%%%%%%%%%%%%%%%%%%%%%%%%%%%%%%%%%%%%
%%%%%%%%%%%%%%%%%%%%%%%%%%%%%%%%%%%%%%%%%%%%%%%%%%%%%%%
%%%%%%%%%%%%%%%%%%%%%%%%%%%%%%%%%%%%%%%%%%%%%%%%%%%%%%%

\section{Generation of the SM predictions}\label{app:genPOWHEG}

In this appendix we document the input parameters and the procedures we followed to obtain the theory predictions and uncertainties for the production of jets in the SM\@. As explained in the main text, we used  the {\tt{POWHEG-BOX}} program~\cite{Alioli:2010xd,Alioli:2010xa}, which provides NLO QCD corrections to jet pair production interfaced with a parton shower.

Events have been generated in parallel runs on a cluster, taking advantage of the parallelization features of the {\tt{POWHEG-BOX-V2}}.
An example of the input parameters used to produce the runs used for this analysis, considering the
parallelization of the runs over 2000 cores, is shown below:
%\\~\\
\vspace{0.1cm}
{\footnotesize{

\begin{minipage}[t]{.4\textwidth}
\begin{verbatim}
numevts 10000
ih1   1      
ih2   1      
ebeam1 3500d0
ebeam2 3500d0
lhans1  260000 
lhans2  260000 
use-old-grid    1 
use-old-ubound  1
ncall1   50000 
itmx1    3     
ncall2   100000
itmx2    1     
foldcsi  5     
\end{verbatim}
\end{minipage}\hfill
\begin{minipage}[t]{.4\textwidth}
\begin{verbatim}
foldy    5     
foldphi  2     
nubound  100000
bornktmin 10d0
bornsuppfact 2500d0
withnegweights 1
doublefsr 1 
hdamp 250
maxseeds 2000
manyseeds  1
parallelstage N
xgriditeration 1
fastbtlbound 1
storeinfo_rwgt 1
\end{verbatim}
\end{minipage}
\\~\\ 
}}

The value of the parameter \texttt{bornsuppfact} is particularly
important to obtain a satisfactory coverage of the phase space and
reasonable statistical uncertainties in the tails of the invariant
mass and jet transverse momentum distributions studied here.  When
producing predictions for a $100$ TeV collider, \texttt{bornsuppfact}
was raised to $25$ TeV\@.

The value of the parameter \texttt{hdamp} affects instead the goodness
of the agreement of {\tt{POWHEG}} predictions with data, especially for the
dijet mass distribution in the central rapidity bin. Despite the fact that
{\tt{POWHEG}} predictions are NLO accurate for inclusive quantities like the
invariant mass of the two hardest jet, or the inclusive jet transverse
momentum, the {\tt{POWHEG}} formula allows for terms beyond NLO to be present
in the final predictions.  The value of the \texttt{hdamp} parameter
ultimately determines the numerical size of these terms. For a more
detailed explanation of the role of \texttt{hdamp}, and its relation
with the fraction of the real-emission matrix element that is
exponentiated by the {\tt{POWHEG}} formula, we refer interested readers
to the original papers~\cite{Alioli:2008tz,Alioli:2009je}.  For this
study, we treated the value of \texttt{hdamp} as a nuisance parameter:
we produced predictions for different values (\texttt{hdamp}$=50, 125,
250, 500, 1000$ and $\infty$ ) and we obtained the best overall fit to
ATLAS data using \texttt{hdamp}$=250$ GeV, which we choose as our
baseline value.

In order to obtain scale and PDF variations, the events in the LHEF
file have been then further processed by the reweighting machinery of
{\tt{POWHEG-BOX-V2}}. This required adding the following lines to the input
card
{\footnotesize{
\begin{verbatim}
rwl_group_events 10000
rwl_file 'reweight.xml'
rwl_add 1
\end{verbatim}}}
and providing a reweighting file in the form
{\footnotesize{
\begin{verbatim}
<initrwgt>
<weightgroup name='Scales'>
<weight id='2'> renscfact=0.5 facscfact=0.5 </weight>
<weight id='3'> renscfact=0.5 facscfact=1.0 </weight>
<weight id='4'> renscfact=1.0 facscfact=0.5 </weight>
<weight id='5'> renscfact=1.0 facscfact=2.0 </weight>
<weight id='6'> renscfact=2.0 facscfact=1.0 </weight>
<weight id='7'> renscfact=2.0 facscfact=2.0 </weight>
</weightgroup>
<weightgroup name='PDFS'>
<weight id='8'> lhapdf=260001 </weight>
<weight id='9'> lhapdf=260002 </weight>
<weight id='10'> lhapdf=260003 </weight>
...
<weight id='106'> lhapdf=260099 </weight>
<weight id='107'> lhapdf=260100 </weight>
</weightgroup>
</initrwgt>
\end{verbatim}}}
The final LHEF file produced after this step contains all the weights 
necessary to evaluate scale and PDF uncertainties.

Finally, we shower each event with two different shower Monte Carlo programs, {\tt{Pythia6}} and {\tt{Pythia8}}. When showering with {\tt{Pythia6}}, we use the Perugia~0 tune~\cite{Skands:2010ak} while showering with {\tt{Pythia8}} is performed with both the Monash 2013~\cite{Skands:2014pea} and the ATLAS A14~\cite{Aad:2014xaa} tunes.
We use the convolution of these results to determine the showering and hadronization uncertainties, which is included in our fit as described in Section~\ref{sec:bounds1}.

%%%%%%%%%%%%%%%%%%%%%%%%%%%%%%%%%%%%%%%%%%%%%%%%%%%%%%%
\section{Double-differential fit}\label{app:ddf}

\begin{figure}[t]
\begin{center}
\includegraphics[width=0.45\textwidth]{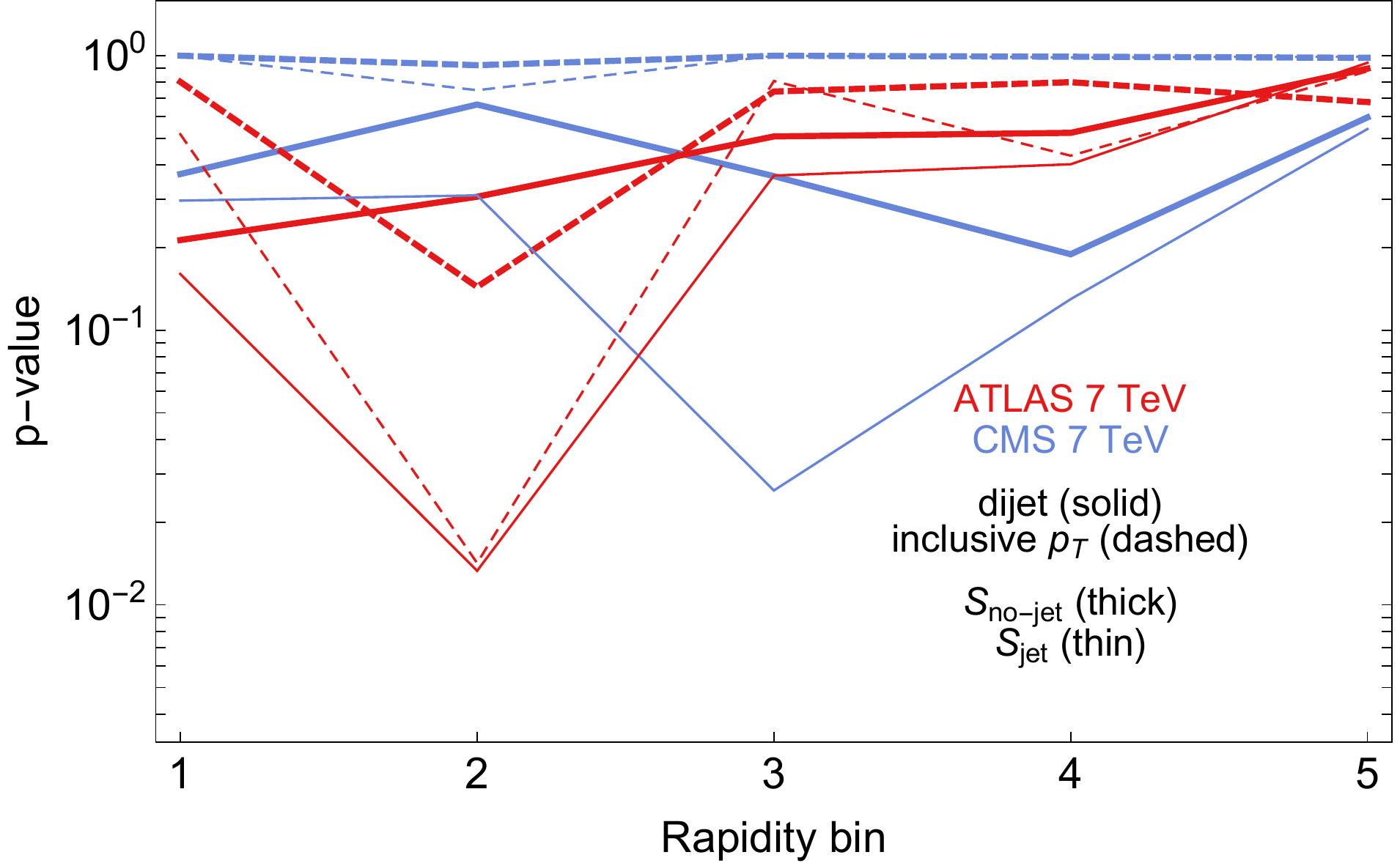}~~~~~\includegraphics[width=0.45\textwidth]{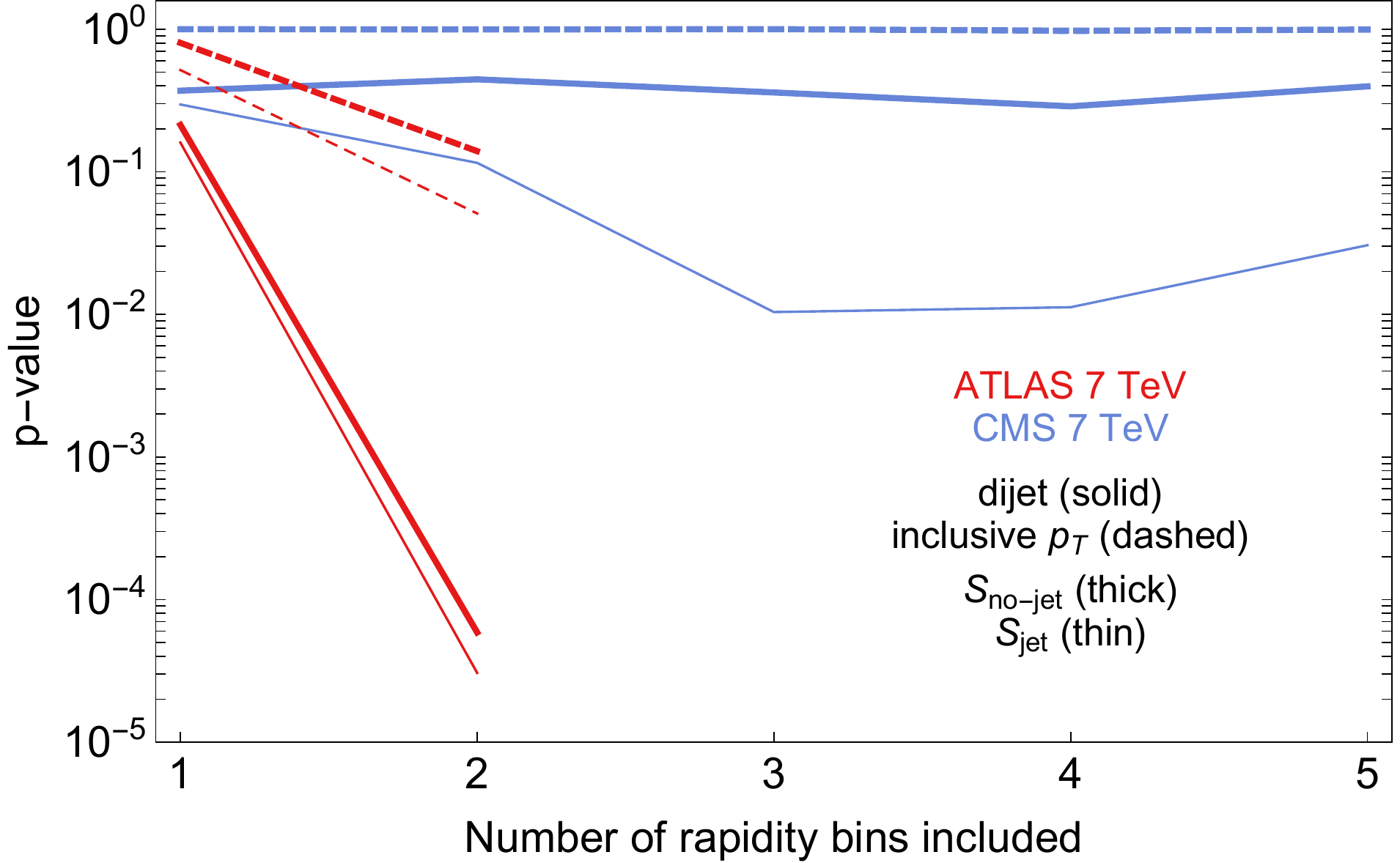}\\ \vspace{1cm}
\includegraphics[width=0.45\textwidth]{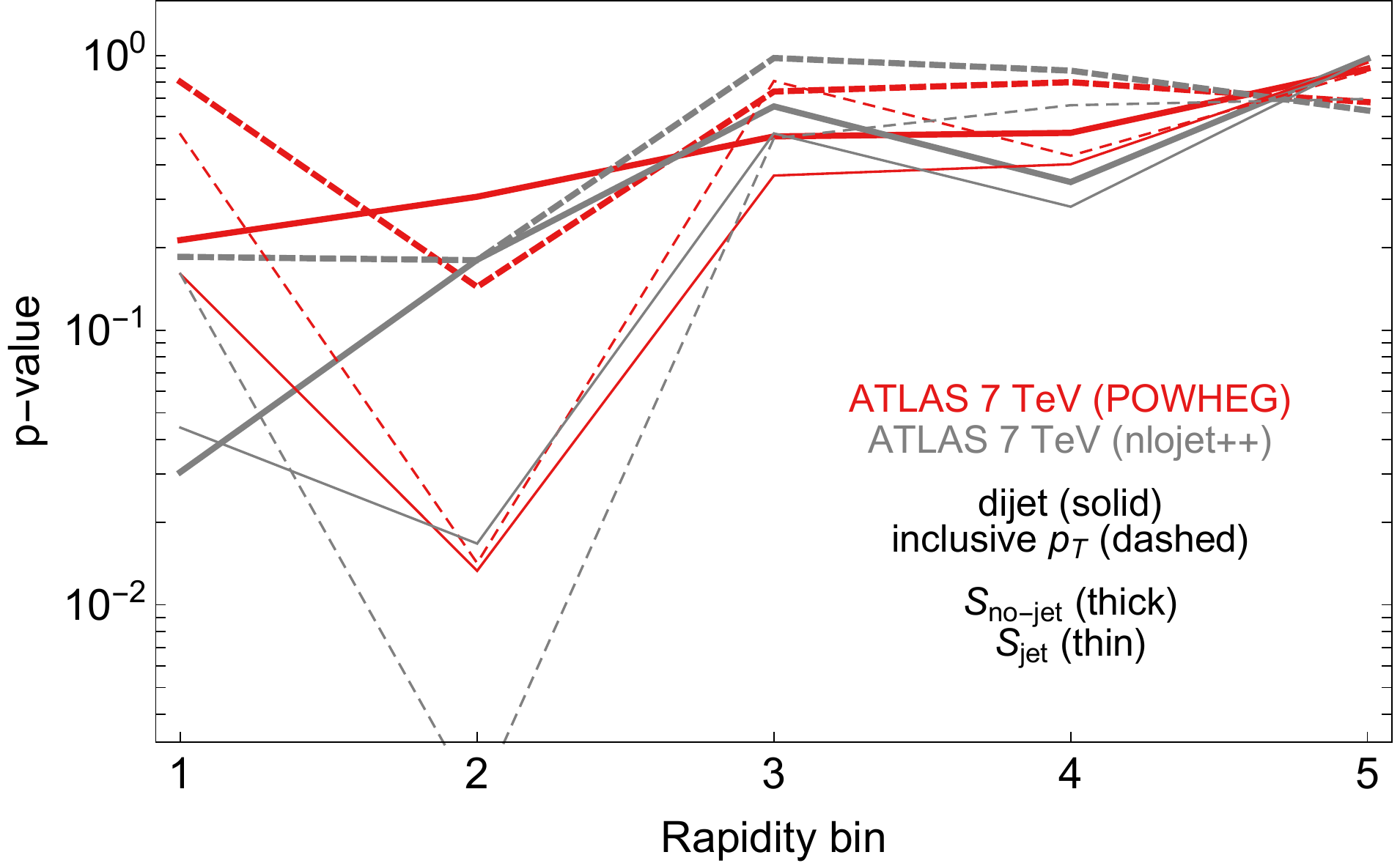}~~~~~\includegraphics[width=0.45\textwidth]{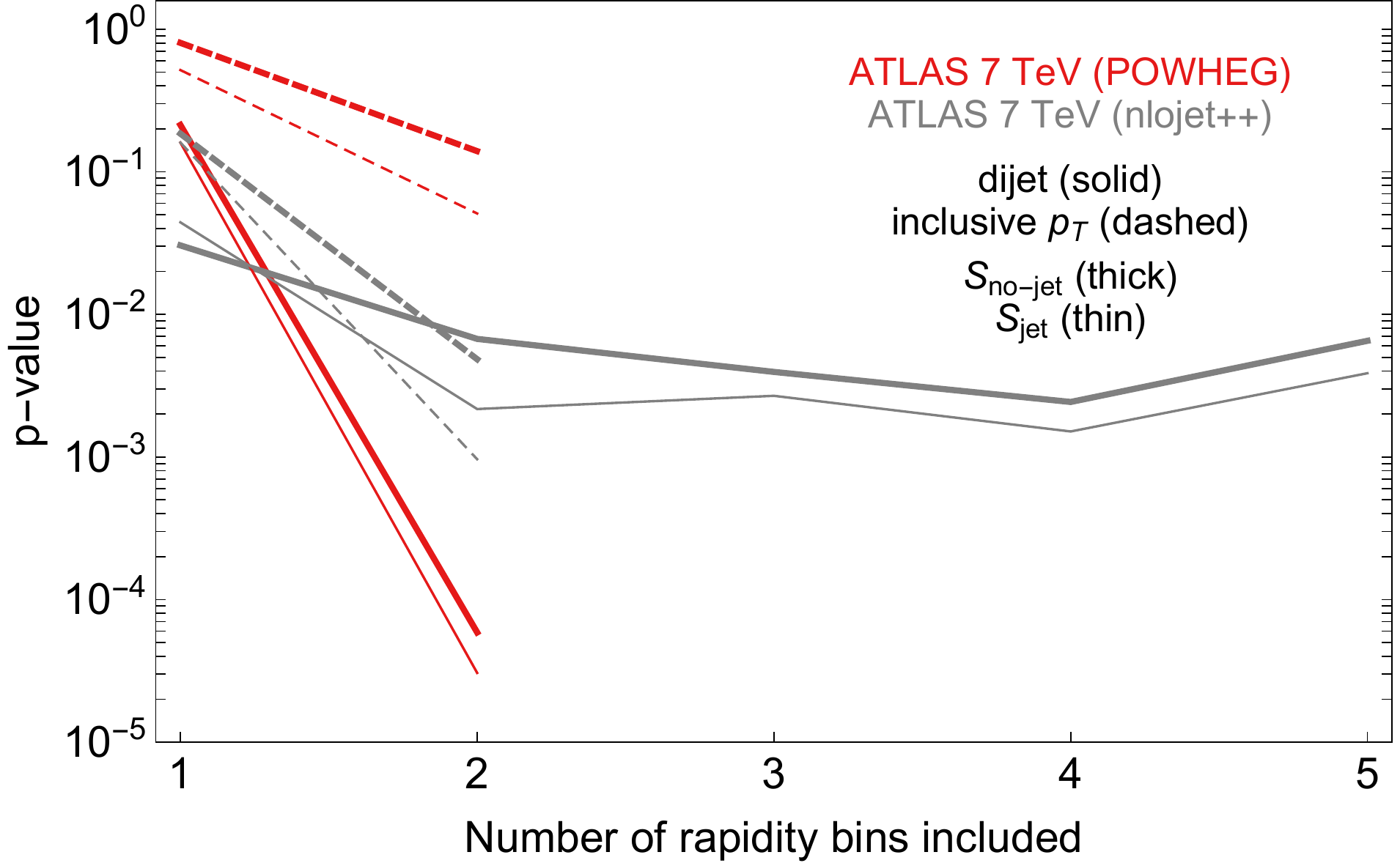}
\end{center}
\vspace{-.3cm}
\caption{ \footnotesize
 p-values measuring the goodness of fit of the SM prediction for the 7\,TeV analyses. They are computed for individual $y$ bins (left panels) and for their combination (right panels). The upper panels show the comparison between ATLAS and CMS for different choices of PDF set, using both dijet and inclusive jet data. The lower panels show only the ATLAS results in which the SM prediction is obtained with two different generators: either {\tt{POWHEG}}, as above, or {\tt{NLOJet++}}. Points that are absent in the combination plots lie outside the range of the plot and correspond to very poor p-values, as explained in the text}
\label{pvalues}
\end{figure}

In order to estimate the quality of our fit to the experimental
measurements, we define a p-value estimator as
 \be \label{eq:pvalue}
{\textrm{p-value}}=1-{\textrm{CDF}}_{\chi^2_N}(\chi^2)  \, ,
\ee 
where
${\textrm{CDF}}_{\chi^2_N}(\chi^2)$ is the cumulative function for the
the chi-squared distribution with $N$ degrees of freedom and $\chi^2$
is evaluated at the SM prediction (corresponding to $\rmZ=0$).  The
p-values for the fit of all individual $y$ bins, and for their
sequential combination, are shown in Fig.~\ref{pvalues}. While
individual bins have in general a sizable p-value, their combination does
not and quickly degrades in the case of ATLAS data, both dijet and
inclusive $p_T$. In particular, the inclusion of 3 or more rapidity
bins brings the p-value below $10^{-5}$ (outside the range of the
plot), indicating a very poor quality of the fit.  For a given search,
the p-value for the two different PDF sets are usually comparable and,
due to the larger uncertainties, $\mathcal S_{\textrm{no-jet}}$ gives a
better fit than $\mathcal S_{\textrm{jet}}$.

%%%%%%%%%%%%%
%%%%%%%%%%%%%
%%%%%%%%%%%%%
%%%%%%%%%%%%%
%%%%%%%%%%%%%
%%%%%%%%%%%%%
%%%%%%%%%%%%%
%%%%%%%%%%%%%

In order to investigate the robustness of the {\tt{POWHEG}} predictions, we
also considered QCD results at fixed NLO, independently obtained by
{\tt{NLOJet++}}~\cite{Nagy:2003tz}. This allows us to estimate the
effect of the terms beyond NLO which are included in the {\tt{POWHEG}}
formula, and the importance of the choice of the renormalization and
factorization scales in the goodness of the fit. For the
{\tt{NLOJet++}}~\cite{Nagy:2003tz} predictions, the renormalization
and factorization scales used the same functional dependence, $\mu =
p_T e^{0.3 y^*}$, adopted in the ATLAS publication~\cite{Aad:2013tea},
which is different from the underlying-Born $p_T$ used in {\tt{POWHEG}}.  The
availability of these fixed order predictions in the form of
{\tt{APPLgrid}}~\cite{Carli:2010rw} grids allowed us to quickly obtain
scale and PDF variations for the observables measured by ATLAS, that
is the dijet invariant mass distribution and the inclusive jet $p_T$
distribution at 7\,TeV\@. Non-perturbative corrections extracted by
experimental collaborations were also available through {\tt{APPLgrid}}
and applied to these results.  The resulting p-values are presented in
the two lower panels of Fig.~\ref{pvalues}, and show a similar behavior
with respect to our {\tt{POWHEG}} predictions, with just a mild
improvement for the dijet doubly differential fit.

%%%%%%%%%%%%%%%%%%%%%%%%%%%%%%%%%%%%%%%%%%%%%%%%%%%%%%%%%%%%%%%%%%%%%%%%%%%%%%%%%%%%%%%%%%%%%%%%%%%%%%%%%%%%%%%%%%%%%%%%%%%%%%%%%%%%%%%%%%%%%%%%%%%%%%%%%%%%%%%%%%%%%%%%%%%%%%%%%%%%%%%%%%%%%%%%%%%%%%%%%%%%%%%%%%%%%%%%

%%%%%%%%%%%%%%%%%%%%%%%%%%%%%%%%%%%%%%%%%%%%%%%%%%%%%%%%%%%%%%%%%%%%%%%%%%%%%%%%%%%%%%%%%%%%%%%%%%%%%%%%%%%%%%%%%%%%%%%%%%%%%%%%%%%%%%%%%%%%%%%%%%%%%%%%%%%%%%%%%%%%%%%%%%%%%%%%%%%%%%%%%%%%%%%%%%%%%%%%%%%%%%%%%%%%%%%%%%%%%%%%%%%%%%%%%%%%%%%%%%%%%%%%%%%%%%%%%%%%%%%%%%%%%%%%%%%%%%%%%%%%%%%%%%%%%%%%%%%%%%%%%%%%%%%%%%%%%%%%%%%%%%%%%%%%%%%%%%%%%%%%%%%%%%%%%%%%%%

%%%%%%%%%%%%%%%%%%%%%%%%%%%%%%%%%%%%%%%%%%%%%%%%%%%%%%%
\section{Validation}\label{app:validation}

Both ATLAS and CMS have used their 7\,TeV data to constrain contact operators affecting jet physics (although not the contact operator that corresponds to $\rmZ$, from Eq.~\ref{Z4f}). In this section we validate our analysis strategy by comparing our results against theirs. Both experimental analysis considered the same four-fermion operator
\be\label{4fermivalidation}
\Delta \mathscr L=\zeta \,\frac{2 \pi}{\Lambda^2}\left({\textstyle{\sum_q}}\bar q_L\gamma^\mu q_L\right)^2 \,\,,
\ee
where $q_L$ stands for left handed quark doublet and $\zeta=\pm1$. The value of $\zeta$ reflect the sign of the SM-NP interference contribution, being negative for $\zeta=1$ and positive for $\zeta=-1$. 

The ATLAS~\cite{Aad:2013tea} analysis uses part of their 7\,TeV data including only dijet invariant masses with $m_{jj} > 1.31$\,TeV and $y^*<0.5$. The limits are presented for destructive interference ($\zeta=1$) only, and for different choices of PDFs. Using NNPDF2.1 and $R=0.6$, the bound on the scale of the operator they obtain is $\Lambda > 7.0$\,TeV\@. Using the same dataset we extract a limit of $9.1$\,TeV\@.

CMS~\cite{Chatrchyan:2013muj} uses data from Ref.~\cite{Chatrchyan:2012bja},  restricting the analysis to inclusive jet production, with $p_T > 501$ GeV and $|\eta|<0.5$. Using CTEQ6.6 PDFs, they obtain a bound of $9.9$\,TeV and $14.3$\,TeV, for $\zeta=+1$ and $\zeta=-1$, respectively. Using the same data, the bounds we obtain are $10$\,TeV and $19$\,TeV, for the same scenarios. Notice however that a precise comparison is hindered by the fact that we are using NNPDF3.0 to extract the bounds.

The results of our validation procedure are summarized in Table~\ref{valid}. Our bounds are  stronger than the experimental one, but always within 30\% of their value. While our calculation of the NP contribution is at leading order in the strong interaction, NLO corrections are included by ATLAS. It is known that their effect is to reduce the bounds on $\Lambda$ by tens of percent~\cite{Gao:2011ha}, so bringing our prediction in agreement with the ATLAS result.

%%%%%%%%%%%%%%%%%%%%%%
%%%%%%%TABLE%%%%%%%%
%%%%%%%%%%%%%%%%%%%%%%
\begin{table*}[ht]
\begin{center}
{\small
\begin{tabular}{c|c|c}  
  &  Experiments & Our Fit  \\ \hline \hline
ATLAS ($\zeta=1$)  & $7.0$\,TeV & $9.2$\,TeV  \\ 
 CMS  ($\zeta=1$)  & $14.3$\,TeV & $19.0$\,TeV \\ 
 CMS  ($\zeta=-1$)     & $9.9$\,TeV & $10.2$\,TeV \\ 
\end{tabular}
}
\end{center}
\caption{\label{valid}\footnotesize Comparison between the bound on the operator of Eq.~\ref{4fermivalidation} presented by the experimental collaborations and the limit extracted with our fitting procedure. 
%The last column is the fractional difference between the experimental results and ours.
}
\end{table*}

%%%%%%%%%%%%%%%%%%%%%%%%%%%%%%%%%%%%%%%%%%%%%%%%%%%%%%%%%%%%%%%%%%%%%%%%%%%%%%%%%%%%%%%%%%%%%%%%%%%%%%%%%%%%%%%%%%%%%%%%%%%%%%%%%%%%%%%%%%%%%%%%%%%%%%%%%%%%%%%%%%%%%%%%%%%%%%%%%%%%%%%%%%%%%%%%%%%%%%%%%%%%%%%%%%%%%%%%%%%%%%%%%%%%%%%%%%%%%%%%%%%%%%%%%%%%%%%%%%%%%%%%%%%%%%%%%%%%%%%%%%%%%%%%%%%%%%%%%%%%%%%%%%%%%%%%%%%%%%%%%%%%%%%%%%%%%%%%%%%%%%%%%%%%%%%%%%%%%%

\pagestyle{plain}
\bibliographystyle{jhep}
\small
\bibliography{biblio}

\end{document}